\documentclass[twoside,a4paper]{article}
\usepackage{pifont}
\usepackage{graphicx}
\usepackage{subfigure}
\usepackage{enumerate}

\usepackage{verbatim}
\usepackage[T1]{fontenc}
\usepackage{amsfonts}
\usepackage{amsmath,amssymb,fancyhdr,amsthm,bm,cite,dsfont}
\usepackage{url}

\newtheorem{thm}{Theorem}[section]
\newtheorem{cor}[thm]{Corollary}
\newtheorem{lem}[thm]{Lemma}
\newtheorem{prop}[thm]{Proposition}

\theoremstyle{definition}
\newtheorem{defn}[thm]{Definition}

\theoremstyle{remark}
\newtheorem{rem}[thm]{Remark}

\def\vf#1#2{\frac{d#1}{d#2}}
\def\fpd#1#2{\frac{\partial #1}{\partial #2}}
\def\ra{\rightarrow}
\def\lra{\leftrightarrow}

\def\R{\mathbb{R}}
\def\N{\mathbb{N}}

\def\xon{\underline{x}}
\def\xbo{\overline{x}}
\def\ph{\phi}
\def\pp{\varphi}

\def\th{\theta}

\def\eps{\epsilon}

\def\epsm{\overline{\eps}}
\def\epsu{\underline{\eps}}
\def\epss{\eps^*}
\def\epst{\eps_-}
\def\epsp{\eps_+}

\def\s{\sigma}
\def\ka{\kappa}
\def\k{\kappa}

\def\a{\alpha}
\def\b{\beta}
\def\g{\gamma}

\def\id{\textrm{id}}
\def\d{\textrm{d}}

\def\be{\begin{equation}}
\def\ee{\end{equation}}
\def\bes{\begin{equation*}}
\def\ees{\end{equation*}}
\def\beq{\begin{eqnarray}}
\def\eeq{\end{eqnarray}}
\def\beqs{\begin{eqnarray*}}
\def\eeqs{\end{eqnarray*}}

\def\F{H}

\def\Sp{S_-}

\def\kaset{{\cal K}}
\def\kasetc{\overline{\kaset}}

\begin{document}

\title{\Large\textbf{Synchronizing pulse-coupled oscillators by constraining the phase response curve}}

\author{\large {Dirk Aeyels and Lode Wylleman}}
%\vspace{0.05cm} \\
%{\small Address}\\
%{\small E-mail: \texttt{dirk.aeyels@ugent.be, lode.wylleman@ugent.be}} }

\date{\today}
\maketitle
\pagestyle{fancy}
\fancyhead{} % clear all header fields
\fancyhead[EC]{D. Aeyels, L. Wylleman}
\fancyhead[EL,OR]{\thepage}
\fancyhead[OC]{Pulse-coupled oscillators with small coupling strength}
\fancyfoot{} % clear all footer fields

\begin{abstract}
We consider networks of weakly pulse-coupled identical oscillators. In an effort to resolve a long-standing problem, we develop an analytic condition on the infinitesimal phase response curve (iPRC) for synchronized dynamic behaviour, extending the well-known result by Mirollo and Strogatz. Oscillators cluster towards synchronization through recurrent absorptions in the case of fully connected networks. We also point out that the same analytic condition guarantees absorption for general networks, and how the condition is extended for non-homogeneous coupling. For a network of neural oscillators of the quadratic-integrate-and-fire type (QIF) we reinterpret our synchronization result into explicit conditions on the QIF-model parameters.
\end{abstract}

\section{Introduction}\label{sec: introduction}

We consider a collection of identical oscillators -- entities that exhibit recurring  behaviour at uniform time intervals. The oscillators are integrated into a network: they are said to be coupled. They impact on one another through some  physical, chemical, biological process or an energy or information carrier,  depending on the nature of the oscillators. The coupling may be sustained in time or impulsive. A prototype of sustained coupling in time is the Kuramoto model \cite{Kuramoto} which has been studied extensively, while impulsive stimuli, although pervasive in science \cite{MauSacSep12}  received  less attention. Here we focus on oscillators interacting through recurring impulsive stimuli from other oscillators.

This arrangement comes with a set of fundamental and basic questions: Will the interaction induce organized behaviour of the oscillators, e.g. synchronization -- oscillators cannot be distinguished one from another? Under what conditions? Is synchronization possible at all, or are less demanding patterns to be expected?

We will focus on synchronization for a network of all-to-all, identically coupled oscillators and formulate conditions such that all oscillators are  moving in step.
We will also pay attention to networks of arbitrarily pulse-coupled oscillators and present results on
synchronization under extra  assumptions.
%  that once oscillators stick together, they remain so and are never  separated by the interaction.

As an illustration of our theory, we discuss synchronized spiking for a network of neural oscillators of the so-called quadratic integrate-and-fire (QIF) type. While synchronization of leaky integrate-and-fire (LIF) oscillators has been studied extensively (see, e.g., \cite{Peskin,MirStr90,Timme1}), it has been argued that quadratic integrate-and-fire (QIF) relaxation oscillators are a more accurate description for some types of neural oscillators (see, e.g., \cite{ErmKop86,HopIzh97,Izhbook}). Nevertheless, results on organized behaviour or synchronization results for the QIF case or other cell types remain heavily underdeveloped~\cite{MauSacSep12}.

Needless to say that synchronization conditions will depend heavily on the dynamics of the oscillators and on the type of interaction. Unfortunately analytical results on synchronization remain elusive except under some crucial monotonicity condition on the infinitesimal phase response curve (iPRC) to be introduced shortly.

In an attempt to formulate and explain  our results it is convenient to provide a mathematical description of our setup as a phase model.
Any real life oscillator can be modeled as a point on a circle, moving with uniform velocity $\omega=\frac{1}{T}$, as will be made explicit in the next section.
Here $1$ stands for the normalized circumference of the circle that models the oscillator orbit, and $T$ is the period: the time it takes for the oscillator to move around its orbit or for the corresponding point to move around the circle. Choosing some preferred point on the circle, one can assign a  phase variable $\varphi$ to the oscillator point. This phase is the {\em normalized  arc length}, measured counterclockwise, from the preferred point to the oscillator point, and is reset from 1 to 0 whenever the oscillator point  passes the preferred point.
This straightforward representation -- the oscillator's {\em phase model} --
not only captures the dynamic behaviour of the oscillator itself but is also a reduced description of the dynamical behaviour in the neighborhood of an attracting oscillator \cite{Winfree}. Notice that the model
denies any information on the amplitude of the oscillator. Nevertheless, it suffices for broad purposes (see, e.g., \cite{Winfree}) and is nonspecific to any particular scientific field.

%The dynamics of the unperturbed oscillator has been reduced  to  a {\em phase oscillator} with a uniform time behaviour, obviously loosing vital information in the process (e.g. concerning the change of the variables in time). The reduction process of the oscillator dynamics to the trivial dynamics on $S^{1}$ can actually be extended to a neighborhood of the oscillator in case it is asymptotically attracting. {\bf [Winfree, ... ]} Each point in this neighborhood is then associated with a phase on the circle such that their behaviour is indistinguishable in the limit for large time.

The uniform periodic behaviour displayed by each oscillator will be affected by the oscillators it is  connected to. We assume this influence
to be {\em impulsive}: each oscillator,  when crossing the preferred point, modifies the connected oscillators' phases by some phase shift. There is the implicit assumption that the impulsive coupling is small enough with respect to the converging properties of the attracting orbit such that the periodic orbit is regained before the next impulsive input.

The phase shift depends on the phase $\varphi$ of the receiving oscillator, on the strength $\epsilon$ of the impulse, and obviously on the dynamics of the oscillator in the neighborhood of the periodic orbit and is captured by the Phase Response Curve taking values  denoted by $\text{PRC}_{\epsilon}(\varphi)$. Recall that oscillators, if not being affected by other oscillators are running free with uniform velocity. In summary, we are  dealing with a hybrid system where each oscillator observes uniform behaviour, resetting its phase quantified by  $\text{PRC}_{\epsilon}(\varphi)$ when other connected oscillators are crossing their preferred positions.

%The phase resetting is described by the Phase Response Curve (PRC), depends in general on the phase and the input from outside (in our case the other oscillators). Since we are considering impulsive stimuli, the PRC will depend on $\varphi$ and the strength $\epsilon$ of the impulse.

The
%$\text{PRC}_{\eps}(.)$
$\text{PRC}_{\eps}$ captures the information of the impulsive interaction between oscillators, and has the important property  that it may be measured by experiment, or determined through simulations or possibly calculated when a mathematical model is available.
Our results will therefore primarily be developed and formulated in terms of the
%$\text{PRC}_{\eps}(.)$
$\text{PRC}_{\eps}$ and more specifically in terms of  its infinitesimal version
% $\text{iPRC}(.)$,
$\text{iPRC}$ a linear approximation of the
%$\text{PRC}_{\eps}(.)$
$\text{PRC}_{\eps}$ for impulsive inputs of small strength $\epsilon$.

For {\em neural}
integrate-and-fire oscillator networks,
the PRC$_\eps$ and $\text{iPRC}$ take a particular form, derived from a phase-state map $f$ (see Sec.\ \ref{subsec: IF}). The assumption that the neural pulse coupling is {\em LIF-like}  amounts to the condition that the $\text{iPRC}$ has a strictly positive derivative everywhere:~\footnote{
%\label{foot:notation}
Equivalently, the PRC$_\eps$ has a strictly positive derivative in some interval $[0,E(\eps)]$, with $0<E(\eps)<1$, see
%\eqref{LIF1} and
\eqref{LIF-2} further below.
%Sec.\ \ref{subsec: LIF}}.
Throughout, we use prime notation to denote the ordinary derivative of a function $q=q(s)$.  If we say that $q$ is continuously differentiable on $[a,b]$ then $q'(a)\in\R$ and $q'(b)\in\R$ are to be interpreted as right and left derivatives, respectively.}
\begin{equation}\label{intro:mirollostrogatz}
\text{iPRC}'(\varphi) >0, \quad \varphi\in [0,1].
\end{equation}
Returning to problems of synchronization (i.e.\ all oscillators sharing the same phase), condition \eqref{intro:mirollostrogatz} revealed itself crucial
in the seminal paper by Mirollo and Strogatz~\cite{MirStr90}, who proved that the equivalent condition $f''(\pp)<0,\,\forall\pp\in[0,1]$ secures synchronization for a totally interconnected network of LIF-like neural oscillators, for arbitrary impulse strength $\epsilon$ and for almost all initial oscillator phases.

%{\bf L: volgende zin nog aanpassen}
A major feature of this paper is that  more general PRC$_\eps$'s and $\text{iPRC}$'s are accepted  than in the case of LIF-like neural pulse coupling.
%, promoting these maps to essential indicators for synchronization, more essential than properties of the potential function in the neural case.}
We dispose  of condition \eqref{intro:mirollostrogatz} and show that for an $\text{iPRC}$ satisfying the weaker condition
%~\footnote{\label{foot: prime notation} To denote the ordinary derivative of a univariate function $q=q(s)$ we will use either differential or prime notation. If we say that $q$ is continuously differentiable on $[a,b]$ then $q'(a)\in\R$ and $q'(b)\in\R$ are to be interpreted as right and left derivatives, respectively.}
\begin{equation}\label{intro:maincondition}
\text{iPRC}'(\varphi)+\text{iPRC}'(1-\varphi) >0, \quad \varphi\in [0,1],
\end{equation}
synchronization is guaranteed, for almost all initial oscillator phases, and for weak pulse coupling, i.e. for impulses of small intensity $\epsilon$.
%belongs to a certain interval $]0,\epsilon_+[$, where $\epsilon_+$ is strictly positive and may depend on the unperturbed oscillator model.

The condition is brought about by exploring the weak coupling assumption. Starting from the uncoupled case ($\epsilon$=0) which corresponds to a straightforward dynamics, the condition \eqref{intro:maincondition}
emerges  in a natural way and is then shown to be effective as a condition for absorption in the small $\epsilon$ case.
We point out that the proof  is such that for  LIF-like  neural models, $\epsilon$ may be chosen  arbitrary large,  recovering the result of \cite{MirStr90}. Just as in \cite{MirStr90}, synchronization is accomplished through a process of successive  absorptions, where oscillators continue absorbing one another until  full synchronization.

Obviously condition \eqref{intro:maincondition} is a weaker requirement on the iPRC for absorption. Notice also that the conditions are conceptually different. While condition \eqref{intro:mirollostrogatz} is of a local nature, condition \eqref{intro:maincondition} is global, reflecting the presence of the circle as the space where the oscillators live on.

Condition \eqref{intro:maincondition} applies to excitatory (i.e., phase-advancing or positive) pulse-coupling. We will assume this type of coupling almost throughout, providing a summary of  analogous results for inhibitory (phase-receding or negative) coupling.

We also point out  that the technique for proving that \eqref{intro:maincondition} is a sufficient condition for synchronization in an all-to-all coupled network, can be adapted and extended, such that  condition \eqref{intro:maincondition} remains unchallenged for {\em absorption} in {\em arbitrarily connected} networks. Summarizing, absorption for weakly pulse-coupled oscillators takes place under reasonable sufficient conditions and is extremely robust: network structure changes  will not impair the absorption phenomenon. As long as the network remains connected, absorption marches on, eventually leading to synchronization if one is willing to accept that clusters of  synchronized oscillators never break up. We also mention how  condition \eqref{intro:maincondition} can be extended for non-homogeneous pulse coupling (with $\text{iPRC}$'s depending on both firing and receiving cells). A precise and convenient mathematical setting to deal with arbitrary connectivity and pulse coupling will be introduced in the follow-up paper \cite{WyllAey14}.

Finally, for a fully and identically coupled network of QIF neural oscillators, we compute an analytic expression for the iPRC; application of \eqref{intro:maincondition} leads then to  sufficient conditions on the QIF-model parameters  for synchronized neural spiking.

\section{Networks of oscillators}\label{sec: networks}

This section is about modeling: First we introduce the oscillator phase model and discuss how its phase may be altered by external influences -- in this paper induced by other oscillators. We recall standard concepts such as the Phase Response Curve (PRC) to represent  phase changes caused by external pulses. We  interconnect the oscillators into a network structure  and model  explicitly the  interaction procedure through pulse-coupling. The synchronization behaviour that we focus on depends crucially on the type of PRC, which we allow to be more general, compared to what is standard in synchronization studies in the literature \cite{MirStr90}. Finally we formulate the dynamics of the overall system in terms of relevant mappings.

\subsection{Oscillator model and oscillator interaction}\label{subsec: oscillator}

%{\bf L: De volgende paragraaf is herschreven}

An oscillator is a structure described by a set of quantifiable variables with a periodic behaviour in time, denoted by $t$: after a positive time $T$, the cycle repeats itself, mapping out a periodic orbit in the set of variables.
To each point on the periodic orbit  a `periodic time' $t^*=t \pmod T \geq 0$ is assigned, being the normalized time it takes to reach that point, starting from a preferred position on the periodic orbit to which the `periodic time' $t^*=0$ is assigned.

The oscillator orbit may be represented by the circle $S^{1}$, with circumference 1. The preferred position on the periodic orbit is identified with a preferred point
%$\O$
on $S^1$, in both cases called the origin and corresponding to the identified `periodic times' $t^{*}=0$ and $t^{*}=T$, while the point on the orbit with $t^*\neq 0$ is identified on $S^1$ with the endpoint of the arc, having length $\pp=\frac{t^*}{T}$ measured counterclockwise from the origin.
%$\O$.
This normalized arc length $\pp\in(0,1)$ is called the {\em phase} of the oscillator (or of the associated point on $S^1$). Assigning the value $\pp=0$ to the origin, the phase is subject to a uniform time evolution $d\pp/dt=\frac{1}{T}$ for $\pp \in [0,1)$. Whenever the oscillator point reaches the origin
%$\O$
in the limit $\pp{\ra} 1$, the phase is reset to $\pp=0$.

The dynamics of the unperturbed oscillator has been reduced  to  a {\em phase oscillator} with a uniform time behaviour, obviously loosing vital information in the process (e.g. concerning the amplitude of the variables and how they change in time). The mapping of the oscillator dynamics to the trivial dynamics on $S^{1}$ can actually be extended to a neighborhood of the oscillator in case it is asymptotically attracting \cite{Winfree}. Each point in this neighborhood is then associated with a phase on $S^{1}$ such that their respective behaviours are  indistinguishable in the limit for large time. Therefore the  {\em phase model}  captures the dynamics
% of the oscillator itself but is also a reduced description of the dynamical behaviour
in the neighborhood of an attracting oscillator.

When an oscillator is subjected to a pulse of strength $\epsilon > 0$ its corresponding phase $\varphi$ {\em advances} to a new $\kappa_{\eps}(\pp)>\pp$ described by  the {\em phase transition curve} $\text{PTC}_{\epsilon}$, denoted as  $\ka_\eps: (0,1) \ra (0,1]$ having properties detailed in  P3 below.  The $\text{PTC}_{\epsilon}$  contains valuable information on the oscillator behaviour and may be derived experimentally, or by simulation, or analytically if a mathematical model is available. We emphasize that $(0,1)$, as the domain  of the PTC$_\eps$,~\footnote{Exclusion of the boundary points 0 and 1 in the domain of the PTC$_\eps$ models the assumption that a firing oscillator does not influence another firing oscillator.} is an open interval of $\R$ where the boundary points $0$ and $1$ are {\em not} identified. Hence, when writing, e.g., $\ka_\eps'(0)$ and $\ka_\eps'(1)$, we mean the (possibly different)  derivatives of the PTC$_\eps$ in 0 and 1, defined as limits of $\ka_\eps'(\pp)$. % (cf.\  footnote \ref{foot:notation}).
The points $\varphi$ of $(0,1)$ do have the interpretation of phases, i.e., normalized arc length measured from the origin. The same remark holds for the $\text{PRC}_{\epsilon}$ and the iPRC introduced below. The boundary point $1$ of the codomain $(0,1]$ of the PTC$_\eps$ indicates that the phase is pulled up to the limit value $1$ by the pulse, after which it is immediately reset to $0$ (see also Sec.\ \ref{subsec: network}).

The $\eps$-dependence of the $\text{PTC}_{\epsilon}$ and the limit case of zero pulse strength  take a central position in our approach.  We  consider a collection $\{\ka_\eps|\eps\in[0,\epsm]\}$, where $\epsm>0$ is some upper bound and
with the following properties:
\begin{enumerate}
\item[P1.] $\ka_0(\pp)=\pp$, for all $\pp\in(0,1)$.
\item[P2.] $\ka_{\epsm}(\pp)=1$, for all $\pp\in(0,1)$.
\item[P3.] For any $\eps\in(0,\epsm)$ the map $\ka_\eps:(0,1)\ra(0,1]$ is phase-advancing
 ($\ka_\eps(\pp)>\pp,\,\forall\pp\in(0,1)$),
 {\em strictly increasing}
 and twice continuously differentiable (i.e., of class $C^2$)
 on an interval $(0,E(\eps))$, with $0<E(\eps)<1$, and satisfies $\ka_\eps(\pp)=1$ for all $\pp\geq E(\eps)$. The map $E:(0,\epsm)\ra(0,1)$ is continuous, strictly decreasing, and has limits $1$ and $0$ for $\eps$ approaching $0$ and $\epsm$, respectively.
\item[P4.] The map
\begin{equation}\label{def-k}
\ka:\quad \kaset\ra (0,1);
\quad (\eps,\pp)\mapsto
\ka_\eps(\pp),\qquad \kaset\equiv
\{(\eps,\pp)\in \R^2|\eps\in(0,\epsm),\;\pp\in(0, E(\eps))\},
\end{equation}
 can be extended to a function $\ka^*:\R^2\ra\R$, of class $C^2$. ~\footnote{ In fact, only the continuity of the partial derivatives
$\fpd{\k^*}{\pp},\,\fpd{^2\k^*}{\pp^2}$ and $\fpd{^2\k^*}{\pp\partial\eps}$ is required for our purposes.}
\end{enumerate}
Property P1 says that there is no phase-jump when there is no pulse. Regarding
property P3, observe that $\ka_\eps$ takes the constant value $1$ on $(E(\eps),1)$, but need not be differentiable at the point $E(\eps)$. We say that an oscillator phase $\varphi\geq E(\eps)$ is pulled up to $1$ when subjected to a pulse of strength $\eps$. This pull-up is referred to as  oscillator or phase {\em absorption} and is a prominent feature of the dynamical behaviour of a network of pulse-coupled oscillators. Because of
property P2, {\em all} phases are absorbed if the pulse strength takes a critical value $\epsm$.
Property P4, finally,
is a mild regularity assumption that will allow for an elegant proof of Theorem \ref{prop: main}, our main result on absorption.
%Notice that ${\cal K}$ is an open convex region of $\R^2$ because the function $E$ decreases.
By the Tietze extension theorem~\cite{Tietze}, P4 is equivalent with the (unique) continuous extendability (by taking limits) of $\ka$ and its partial derivatives up to second order to the closure
\be\label{Kc}
\kasetc=\{(\eps,\pp)\in \R^2|\eps\in[0,\epsm],\;\pp\in[0, E(\eps)]\}
\ee
of $\kaset$.
All requirements P1-P4 are natural, in the sense that they are met by neural
integrate-and-fire oscillator models, to be introduced soon, for which, moreover, $\ka_{\eps}$ always has different left and right derivatives at $E(\eps)$ (see Sec.\ \ref{subsec: oscillator}).

An equivalent characterization of the response of the phase under external impulses  is by means of the {\em phase response curve} $\text{PRC}_{\epsilon}$ denoted as $z_\eps:(0,1) \ra (0,1)$ and defined by
%as the change of phase
$z_\eps(\pp)=\ka_\eps(\pp)-\pp$. It follows that $z_0(\pp)=0$,  while for $0<\eps \leq \epsm$ the quantity $z_\eps(\pp)$ is the phase advance of an oscillator at phase $\pp$ caused by an $\eps$-pulse. The map $\ka$ defined in \eqref{def-k}, and the map $z$
%also defined on $\kasetc$
 by $z(\eps,\pp)=z_\eps(\pp)=\ka(\eps,\pp)-\pp$, are referred to as the PTC and the PRC, respectively.

A major last concept in the development of our work is the {\em infinitesimal phase response curve} (iPRC) $Z: (0,1) \ra \R$ defined by
\begin{equation}\label{Z-prop}
Z(\pp)=\fpd{z}{\eps}(0,\pp)=\fpd{\ka}{\eps}(0,\pp).
\end{equation}
%{\bf For any fixed $\pp\in(0,1)$ the partial derivative $\fpd{\ka}{\eps}(0,\pp)$ is a limit, the boundary point $(0,\pp)$ of  ${\cal K}$ being approached by $(\eps,\pp)$ from the right.}
%Since $\kaset$ is convex the partial derivative $\fpd{\ka}{\eps}(0,\pp)$ makes sense because $\kaset$ is convex.
%For any $\pp\in(0,1)$
The iPRC represents
a linear approximation of $z$ and $\kappa$ for pulses of small strength $\eps$:
\be\label{Z-by-z}
z(\eps,\pp)=\ka(\eps,\pp)-\pp=\eps Z(\pp) + \eps H(\eps,\pp),\quad \lim_{\eps\ra 0} H(\eps,\pp)=0.
\ee
The iPRC is independent of the pulse strength $\eps$ and our main results will be formulated in terms of the iPRC.

All along we have considered  phase-advancing or excitatory pulses of positive (or zero) strength $\eps$. When  $\eps <0$ we are dealing with inhibitory pulses and {\em receding} phase ($\ka_\eps(\pp)<\pp$). In this case a collection $\{\ka_\eps|\eps\in [\epsu,0]\}$ is considered, where $\epsm<0$ is some lower bound and  $\ka_\eps:(0,1)\ra[0,1)$ is a PTC$_{\eps}$. The corresponding PTC $\ka$ is defined by the analogue of \eqref{def-k} and satisfies  properties P1-P4, where $\epsm$ is replaced by $\epsu$, $(0,\epsm)$ by $(\epsu,0)$, and $(0,E(\eps))$ by $(E(\eps),1)$, $\ka_\eps(\pp)=0$ for all $\pp\leq E(\eps)$, and $E$ has limits $1$ and $0$ if $\eps$ approaches $\epsu$ and $0$, respectively.
The results in this paper are laid out for positive $\eps$, but may be adapted to the case of negative $\eps$ (see Sec.\ \ref{subsection: inhibitory}).\\

\vspace{.5cm}

\begin{figure}[h!]
    \begin{center}
        \subfigure[]{
            \label{fig: exc-a}
            \includegraphics[width=0.4\linewidth]{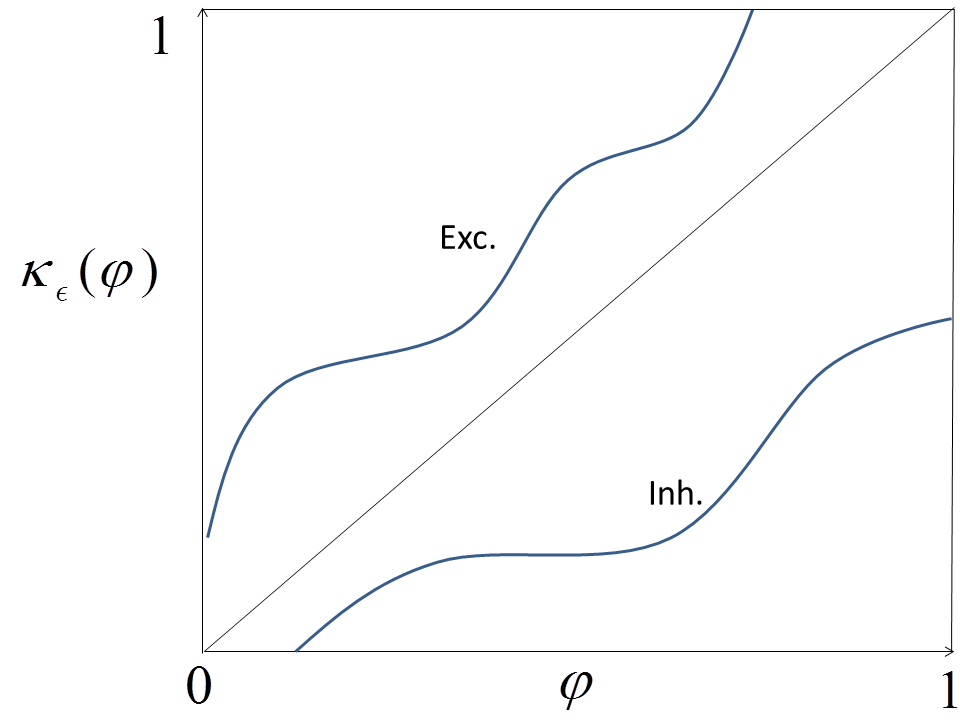}
        }
        \subfigure[]{
           \label{fig: exc-b}
           \includegraphics[width=0.4\linewidth]{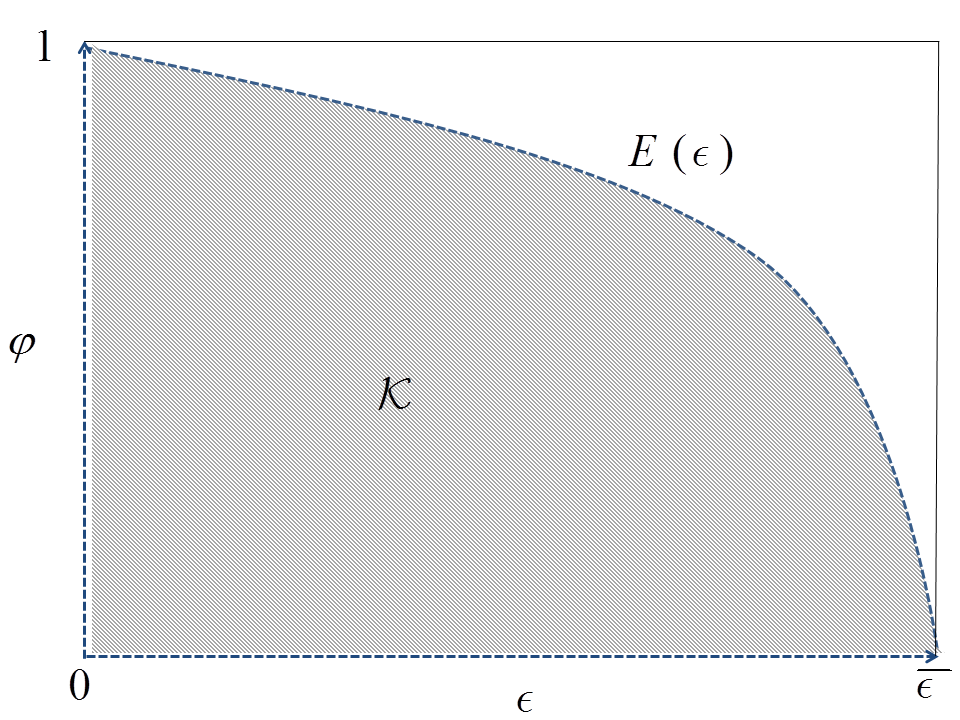}
        }
    \end{center}
    \caption{(a) Examples of an excitatory (Exc.) and inhibitory (Inh.) PTC$_\eps$. (b) Example of a function $E=E(\eps)$ and the corresponding region $\kaset$ in the excitatory case.}
    \label{fig: exc}
\end{figure}

%{\bf Figure ****} shows PTC$_\eps$'s in the excitatory and inhibitory cases.

\subsection{Network model}\label{subsec: network}

We consider {\em networks}  of pulse-coupled  oscillators with phases $\pp_\a$. Here and below, greek letters $\alpha,\beta,\ldots$ are elements of an index set $\{1,\ldots,N\}$ labeling the oscillators.  An {\em oscillator cluster} or {\em synchronous group}, ${\cal G}$, is a maximal set of oscillators with equal phases: $\pp_\a=\pp_{\cal G}$ for all $\a\in{\cal G}$, where $\pp_{\cal G}$ corresponds to the phase, common to the oscillators belonging to ${\cal G}$. We will study networks with the following characteristics:
%Each cell is represented by a phase variable $\ph_\alpha$ on $S^1$.
\begin{enumerate}
\item All oscillators have {\em identical unperturbed dynamics}:  they have a common oscillation period $T$ and phase dynamics
%{\bf het mod model nog vergelijken met de kappa functie en intervals}
\begin{equation}\label{phasedyn}
    d\pp_\alpha/dt=1/T,\quad\forall \pp_\a\in[0,1).
\end{equation}
\item Phase $\pp_\alpha=1$ corresponds to a particular position of the oscillator. When an oscillator $\alpha$ reaches threshold ($\pp_\alpha=1$) it `fires', sending a pulse to all other oscillators $\beta\neq\alpha$ %in case each oscillator is connected to all other oscillators
    ({\em all-to-all} or {\em global pulse coupling}). %The case of not fully coupled networks will also be considered.
\item The parameter $\eps\in[\epsu,\epsm],\,\epsu<0<\epsm$ models the pulse strength and is independent of the firing oscillator $\alpha$ and receiving oscillator $\beta$. If the oscillators of a cluster ${\cal G}$ fire (in unison), the pulse strength experienced by any oscillator not belonging to ${\cal G}$
%, but connected to at least one oscillator of the cluster
remains unchanged and thus equal to $\eps$ ({\em non-additive pulse strength}). The pulse coupling is {\em (purely) excitatory} if $\eps>0$ and {\em inhibitory} if $\eps<0$. In this paper the main focus is on the excitatory case.
\item The phases of the oscillators $\alpha$ of a firing cluster ${\cal G}$, and the phases of the receiving oscillators $\beta$, instantaneously change
from $\pp_\alpha=\pp_{\cal G}=1$ to $0$ and
from $\pp_\beta$ to $\ka_\eps(\pp_\beta)$. The function $\ka_\eps$ is a common PTC$_\eps$, independent of the firing cluster and receiving oscillator $\beta$ ({\em identical} or {\em homogeneous pulse coupling}).
%respectively ({\em zero reset time} (see section \ref{subsec: oscillator}), {\em zero phase transition time} and {\em zero pulse travelling (or interaction) time}).
\item When an oscillator cluster ${\cal G}$ fires and $\ka_\eps(\pp_\beta)=1$ (excitatory coupling) or $\ka_\eps(\pp_\beta)=0$ (inhibitory coupling) for one or more receiving oscillators $\beta$, they are {\em instantaneously absorbed} into ${\cal G}$, their phases instantaneously changing to $0$ along with $\pp_{\cal G}$.
\end{enumerate}

Due to \eqref{phasedyn}, phase differences between  oscillators remain unchanged between firings. The absorption phenomenon introduced in assumption 5 is the pathway to synchronization. Oscillators firing may lead to absorption and as such generate larger clusters. Consecutive absorptions of oscillator clusters may finally result in global synchronization.
% In this respect
%, and when there is all-to-all coupling,

Assumptions 1 and 2, along with the independence on the receiving oscillator of the pulse strength and the PTC$_\eps$, imply that a firing oscillator cluster and the oscillators absorbed by this firing event form a new, larger cluster of oscillators with identical phases from that moment on. Moreover, assumption 3 on the pulse strength and the independence of the PTC$_\eps$ on the firing cluster imply equivalence between clusters and single oscillators.

\subsection{Integrate-and-fire oscillators and neural networks}\label{subsec: IF}

An oscillator is said to be of the {\em integrate-and-fire (IF)} type~\cite{Lapicque,Abbott} if it
is represented by a scalar, time-dependent {\em state} variable $x=x(t)\in \R$, continuously repeating the following cycle:
it increases monotonically (`integrates') from a lower threshold value $\xon$ to an upper threshold value $\xbo$, then `fires', after which it is instantaneously reset to $\xon$, completing the cycle and as such implementing an oscillator.~\footnote{In fact, every oscillator as defined in Sec.\ \ref{subsec: oscillator} can be naturally interpreted as an IF oscillator, by taking for $x$ the arc length along the oscillator orbit, which is reset after completing  the cycle.} The dynamics of the integrating stage
%, $x\in[\xon,\xbo)$,
is modeled by
% a real valued, {\bf continuously differentiable} function $F:[\xon,\xbo]\ra\R_{>0}$, strictly positive on the interval $[\xon,\xbo)$:
\begin{equation}\label{statedyn}
x'=F(x),\qquad \forall x\in[\xon,\xbo).
\end{equation}
%{\bf iets zeggen over gedrag in xboven}
Here the map $F$ is strictly positive and continuously differentiable on its domain $[\xon,\xbo]$.
The triple $(F,\xon,\xbo)$ will be called an {\em IF oscillator model}.
The time $\Delta t(x)$ to go from $\xon$ to $x$ along a solution of \eqref{statedyn}, respectively to $\xbo$ (corresponding to the {\em (oscillation) period} $T$), are given by%time that $x$ needs to climb once from $\xon$ to $\xbo$ is thus
\begin{equation}\label{period}
\Delta t(x)=\int_{\xon}^{x} \frac{ds}{F(s)},\qquad T=\Delta t(\xbo).
\end{equation}
Dividing $\Delta t(x)$ by the period yields a strictly increasing bijection $g$ between states $x\in[\xon,\xbo]$ and {\em phases} $\pp\in[0,1]$, referred to as the {\em state-phase map}, with inverse $f\equiv g^{-1}$ :
\begin{equation}\label{state-phase}
\pp\equiv \frac{\Delta t(x)}{T}\equiv g(x)\quad\leftrightarrow\quad x=f(\pp).
\end{equation}
%The instantaneous character of the state reset $\xbo\mapsto \xon$ is an approximation, expressing that the oscillation period $T$ is much larger than the reset or `relaxation' time, the latter being modelled as zero.
%This
The instantaneous state reset $\xbo\mapsto \xon$ is equivalent to the instantaneous phase reset $1\mapsto 0$. Identifying the endpoints of the interval $[0,1]$ to obtain the unit circle $S^1$, the unperturbed state dynamics of the IF oscillator is faithfully represented by the trivial time evolution $d\pp/dt=1/T$ of an angular phase variable  $\pp$ on $S^1$ (as implied by \eqref{statedyn} and \eqref{state-phase}).

From \eqref{state-phase} one obtains
\be\label{g-deriv_is_F}
\forall x\in[\xon,\xbo]:\;\;g'(x)=\frac{1}{T.F(x)}\quad\lra\quad \forall\pp\in[0,1]:\;\;f'(\pp)=T.F(f(\pp)).
\ee

%**** Inverse question: if one has a strictly positive function, is it then the iPRC for {\em some} integrate-and-fire model? ****
%The iPRC takes a central place in the study of coupled IF oscillators in general, and in the present work in particular.
%The iPRC concept takes a central place in the study of coupled IF oscillators in general, and in the present work in particular.

A network of IF-oscillators is called {\em neural} if the oscillators interact through the following specific (additive) type of pulse-coupling. The effect on an IF oscillator of a {\em pulse} of strength $\eps$ (provoked  by another `firing' oscillator) is to map the state $x$ to $\min(x+\eps,\xbo)$ if $\eps\geq 0$ (positive or {\em excitatory} pulse), and to $\max(x+\eps,\xon)$ if $\eps\leq 0$ (negative or {\em inhibitory} pulse).  As before, we will restrict our discussion to excitatory pulse-coupling.
%nonnegative $\eps$.
%Evidently, if $\eps=0$ then nothing happens and we are back in the unperturbed case. Also, if
Notice that if $\eps\geq \epsm$, where
\be\label{def-epsm-IF}
\epsm=\xbo-\xon
\ee
then any state $x$ is trivially mapped to $\xbo$ and its corresponding phase to $1$ (all cells are absorbed by the firing oscillator). Therefore we  consider pulses $\eps$ in the interval $[0,\epsm]$. In terms of phases, the effect of such a pulse  is captured by the corresponding PTC$_\eps=\ka_\eps: (0,1) \ra (0,1]$, for which we have the following explicit formula in terms of the mathematical model of the IF-oscillator:~\footnote{In the inhibitory case $\epsu=-\epsm<\eps<0$, $x$ maps to $\max(x+\eps,\xon)$, $\ka_\eps(\pp)$ equals $g(f(\pp)+\eps)$ if $\pp> g(\xon-\eps)$ and $0$ else.}
\begin{equation}\label{def-keps}
\ka_\eps(\pp)=g(\min(f(\pp)+\eps,\xbo))=
\begin{cases}
g(f(\pp)+\eps),& \pp< g(\xbo-\eps);\\
1,& \pp\geq g(\xbo-\eps).
\end{cases}
\end{equation}
%if the pulse is excitatory ($\eps\in(0,\epsm]$).
% and by
%\begin{equation}\label{def-keps}
%\ka_\eps(\pp)=g(\max(f(\pp)+\eps,\xon))=
%\begin{cases}
%g(f(\pp)+\eps),& \pp> g(\xon-\eps);\\
%0,& \pp\leq g(\xon-\eps)
%\end{cases}
%\end{equation}
%if the pulse is inhibitory ($\eps\in[-\epsm,0]$).
%Notice that there is no interaction if the pulse has zero strength, i.e., $\ka_0(\pp)=\pp,\,\forall \pp \in(0,1)$.
%Since $f$ and $g$ are strictly increasing, so are the maps $\ka_\eps$, for all $\eps\geq 0$.
A collection
%$\{\ka_\eps|\eps\in[0,\epsm]\}$
$\{\ka_\eps|\eps\in[0,\epsm]\}$
of these maps has the four PTC properties P1-P4 outlined in Sec.\ \ref{subsec: oscillator}. Properties P1 and P2 are obviously satisfied.
%, with $\epsm=\xbo-\xon$.
Property P3 is satisfied, with $E(\eps)=g(\xbo-\eps)$, because of the strictly increasing character of $f$ and $g$, the assumptions on $F$, and
\be\label{ka-eps'}
\ka_\eps'(\pp)=g'(f(\pp)+\eps)f'(\pp)=\frac{g'(f(\pp)+\eps)}{g'(f(\pp))}=\frac{F(f(\pp))}{F(f(\pp)+\eps)},\quad \forall\pp\in[0,g(\xbo-\eps)].
\ee
Notice that the left derivative of $\ka_{\eps}$ at $g(\xbo-\eps)$, viz.\ $F(\xbo-\eps)/F(\xbo)$, never equals the right derivative value $0$.
Finally, consider the map $\ka$ defined in \eqref{def-k}. By \eqref{g-deriv_is_F} and the assumptions on $F$,
%$0<F(x)<+\infty$ on $[\xon,\xbo]$, and by the continuity of $F$ and $F'$ on $[\xon,\xbo]$,
we can extend $g$ to a twice continuously differentiable, strictly increasing bijection $g^*:\R\ra\R$, with inverse $f^*$. Given a pair $(f^*,g^*)$, the map
$\ka^*:\R^2\ra\R$ defined by $\ka^*(\eps,\pp)=g^*(f^*(\pp)+\eps)$ is a continuous extension of $\ka$ having continuous partial derivatives up to second order, such that also property P4 is satisfied.

The iPRC $Z$ as defined in \eqref{Z-prop}, based on \eqref{def-k} and the explicit formula \eqref{def-keps} is given by
\begin{equation}\label{def-iPRC}
Z(\pp)=\fpd{\ka}{\eps}(0,\pp)= g'(f(\pp)) \equiv\vf{\pp}{x}(f(\pp))= \frac{1}{f'(\pp)}=\frac{1}{TF(f(\pp))}.
\end{equation}
%\begin{equation}\label{def-iPRC}
%Z(\pp)\equiv \vf{\pp}{x}(f(\pp))\equiv g'(f(\pp)) =\frac{1}{f'(\pp)}=\frac{1}{TF(f(\pp))}.
%\end{equation}
%{\bf moet ik hier iets zeggen over definitiegebied?}
Notice that $Z$ returns the quotient of the infinitesimal change of phase caused by an infinitesimal change of state, as a function  of the phase.
%Since $g$ (or equivalently $f$) is strictly increasing,
By strict positivity of $F$, the iPRC is also a strictly positive function. Its derivative features prominently in our sufficient condition for synchronization and is computed for all $\pp\in[0,1]$ as
\be\label{Z-cond-alt}
Z'(\pp)=\left(\frac{1}{f'}\right)'(\pp)=-\frac{f''}{f'^2}(\pp)=(g'\circ f)'(\pp)=\frac{g''}{g'}(f(\pp))=-\frac{F'}{F}(f(\pp)).
\ee

A neural network of pulse-coupled IF oscillators, with $(F,\xon,\xbo)$ as the underlying IF oscillator model and
%meeting the requirements of Sec.\ \ref{subsec: network} for a
pulse strength $\eps>0$, will be denoted as $(F,\xon,\xbo,\eps)$. The  PTC$_\eps$ associated to such a neural network (i.e., a $\ka_\eps$ of the form \eqref{def-keps}, with $f$ and $g$ defined by \eqref{period}-\eqref{state-phase}), or a PTC $\ka$ representing a collection of such PTC$_\eps$'s, is also called
{\em neural}. %A collection of such neural networks, where $\eps\in[0,\eps_0]$ and $0<\eps_0\leq \epsm=\xbo-\xon$, will be symbolized by $(F,\xon,\xbo,\eps_0)$.

\subsection{LIF-like oscillator networks}\label{subsec: LIF}

In neuroscience, the {\em leaky integrate-and-fire (LIF)} neural network model is a popular simplification of the widely adopted Hodgkin-Huxley model~\cite{HodgHux}. Peskin used a similar model for the cardiac pacemaker~\cite{Peskin}. In both cases an {\em LIF oscillator} is represented by a voltage-like state-variable $x=x(t)$ subject to \eqref{statedyn}, with
\bes
F(x)=S-\gamma x,\qquad S>\gamma>0,\qquad \xon=0,\quad \xbo=1.
\ees
By \eqref{period}, \eqref{state-phase} and \eqref{def-iPRC} we get
\beqs
&&T=\frac{1}{\gamma}\ln\left(\frac{S}{S-\gamma}\right),\quad g(x)=\frac{1}{\gamma T}\ln\left(\frac{S}{S-\gamma x}\right),\quad f(\pp)=\frac{S}{\gamma}(1-e^{-\gamma T\pp})=\frac{1-e^{-\gamma T\pp}}{1-e^{-\gamma T}},\nonumber\\
&&Z(\pp)=\frac{1}{ST}e^{\gamma T\pp},\quad Z'(\pp)=\frac{\gamma}{S}e^{\gamma T\pp},\quad Z''(\pp)=\frac{\gamma^2 T}{S}e^{\gamma T\pp}.
\eeqs
Since $Z'(\pp)>0,\,Z''(\pp)>0,\,\forall\pp$ synchronization occurs for {\em all} initial conditions, except for the unique equilibrium point of the firing map if $\eps<1/n$, a result proved by using contraction techniques~\cite{MauSep08}.

For an {\em arbitrary}  neural network model $(F,\xon,\xbo,\eps)$, \eqref{Z-cond-alt} implies that the following conditions are equivalent:
\beq
\label{LIF1}&&(a)\quad\forall \pp\in[0,1]:\;Z'(\pp)>0;\qquad (b)\quad\forall \pp\in[0,1]:\;f''(\pp)<0;\\
\label{LIF2}&&(c)\quad\forall x\in[\xon,\xbo]:\;F'(x)<0;\qquad (d)\quad\forall x\in[\xon,\xbo]:\;g''(x)<0.
%\label{LIF1}&&(a)\quad\forall \pp\in[0,1]:\;Z'(\pp)>0;\\
%\label{LIF2}&&(b)\quad\forall \pp\in[0,1]:\;f''(\pp)<0;\\
%\label{LIF3}&&(c)\quad\forall x\in[\xon,\xbo]:\;F'(x)<0\;\\
%\label{LIF4}&&(d)\quad\forall x\in[\xon,\xbo]:\;g''(x)<0.
%\forall \pp\in[0,1]:\;Z'(\pp)>0\;\lra\;f''(\pp)<0  \qquad\lra\qquad \forall x\in[\xon,\xbo]:\;F'(x)<0\; \lra\; g''(x)>0.
\eeq
The condition $f''(\pp)<0,\,\forall\pp\in[0,1]$ (i.e., a {\em concave down} phase-state map) featured explicitly in the paper by Mirollo and Strogatz~\cite{MirStr90} to obtain synchronization for almost all initial conditions. By \eqref{ka-eps'}, \eqref{LIF1} implies, for any $\eps\in(0,\epsm=\xbo-\xon)$,
\be\label{LIF-2}
%\forall\eps\in(0,\epsm):\quad
\ka_\eps'(\pp)>1,\quad\forall \pp\in[0,g(\xbo-\eps)]\qquad\lra\qquad z_\eps'(\pp)>0,\quad\forall \pp\in[0,g(\xbo-\eps)].
\ee
%i.e., the PRC$_\eps$ is strictly increasing.

Consider a network of pulse-coupled oscillators for which the corresponding phase transition curve $\ka_\eps$ is not necessarily of the neural form \eqref{def-keps}.
%satisfying the requirements of Secs.\ \ref{subsec: oscillator} and \ref{subsec: network} and with pulse strength $\eps$. The corresponding
The PTC properties P3 and P4 in Sec \ref{subsec: oscillator} imply that $\ka'_\eps(\pp)\geq 0$ for any $\eps\in(0,\epsm)$ and $\pp\in [0,E(\eps)]$. Then the first part of \eqref{LIF1}, and \eqref{LIF-2} suggest to introduce the following definition:

\begin{defn}\label{definition: LIF} If $\ka_\eps$ satisfies the stronger restriction $\ka'_\eps(\pp)>1$ on $[0,E(\eps)]$
%~\footnote{Referring to footnote \ref{foot: prime notation}, this definition requires in particular that $\ka'_\eps(0)>1$ and $\ka'_\eps(E(\eps))>1$, whereas $\ka'_\eps(0)=0$ or $\ka'_\eps(E(\eps))=0$ are allowed for a general PTC$_\eps$.}
then
%the network and the PTC$\eps$ are
it is called {\em LIF-like}.
%A PTC $\ka$ is called LIF-like if $Z'(\pp)>0$ for all $\pp\in[0,1]$, the iPRC $Z$ defined by \eqref{Z-prop}.
An iPRC $Z$, associated to a PTC $\ka$ by \eqref{Z-prop}, is called {\em LIF-like} if $Z'(\pp)>0$ for all $\pp\in[0,1]$.
\end{defn}

\begin{prop}\label{prop: LIF-like} If the iPRC $Z$ associated to a PTC $\ka$ is LIF-like then there exists a maximal real number $\epsp>0$ such that for any $\eps\in(0,\epsp]$ the corresponding phase transition curve $\ka_\eps$ is LIF-like. If $\ka$ is neural
%stands for a collection $\{\ka_\eps|\eps\in[0,\epsm]\}$ modeling {\em neural} networks,
then $\epsp=\epsm$, i.e., {\em any} corresponding $\ka_\eps$ is LIF-like.
\end{prop}

\begin{proof} By virtue of PTC property P4 (see Sec.\ \ref{subsec: oscillator}) there exists a continuous extension $\ka^*$ of $\ka$ to $\R^2$, with continuous partial derivatives up to second order, such that
\be\label{Schwarz}
K_{\pp\eps}\equiv \fpd{}{\eps}\left(\fpd{\ka^*}{\pp}\right)=\fpd{}{\pp}\left(\fpd{\ka^*}{\eps}\right)
\ee
%$\partial^2\ka^*/\partial\eps\partial\pp=\partial^2\ka^*/\partial\pp\partial\eps$
by Schwarz's rule.
%We denote the restrictions of $\partial\ka^*/\partial\pp$ and $\partial^2\ka^*/\partial\pp\partial\eps$
%to $\kasetc$ $\kasetc=\bigcup_{\eps\in[0,\epsm]}{\eps}\times [0,E(\eps)]$  by $\ka^*_\pp$ and $\ka^*_{\pp\eps}$, respectively.
%Using $\partial^2\ka^*/\partial\eps\partial\pp=\partial^2\ka^*/\partial\pp\partial\eps$~\footnote{The criteria for applying the Schwarz theorem to $\ka$ are fulfilled by property P4.}, and
Definition \eqref{def-iPRC} and our assumption that the iPRC is LIF-like give
\be\label{ZZ}
K_{\pp\eps}(0,\pp)=\fpd{}{\pp}\left(\fpd{\ka^*}{\eps}\right)(0,\pp)=\fpd{}{\pp}\left(\fpd{\ka^*}{\eps}(0,\pp)\right) =Z'(\pp)>0,\qquad \forall \pp\in[0,1].
\ee
%$\ka^*_{\pp\eps}(0,\pp)=Z'(\pp)>0$ for all $\pp\in[0,1]$.
Since $K_{\pp\eps}$ is continuous the set $K^{-1}_{\pp\eps}((-\infty,0])$ is closed. Hence, its intersection $\Sp$ with the compact set $\kasetc$ (see \eqref{Kc}) is compact. If $\Sp\neq\emptyset$ the continuous projection map $(\eps,\pp)\mapsto\eps$ reaches a minimum $\epss$ on $\Sp$, with $\epss>0$  because of \eqref{ZZ}.
%which is strictly positive by \eqref{ZZ}
%(implying $\{0\}\times[0,1]\nsubseteq\Sp$).
%(implying $\{0\}\times[0,1]\subset\kasetc\setminus\Sp$).
If $\Sp=\emptyset$ then we put $\epss=\epsm$. It follows that $K_{\pp\eps}$ is strictly positive on $[0,\epss)\times\,[0,1]$.
%~\footnote{Since $(-\infty,0]$ is closed and $\ka^{*}_{\pp\eps}$ is continuous, the set $\ka^{*-1}_{\pp\eps}((-\infty,0])$ is the intersection of a closed set with the compact domain $\kasetc$ of $\ka^{*}_{\pp\eps}$, and thus compact.}
%Denote $\ka^*_\pp$ and $\ka^*_{\pp\eps}$ for $\partial\ka^*/\partial\pp$ and $\partial^2\ka^*/\partial\pp\partial\eps$.
%for fixed $\pp\in[0,1]$ we have $\ka^*_{\pp\eps}(0,\pp)=Z'(\pp)>0$.~\footnote{Notice that the partial derivative $\ka_{\pp\eps}(0,\pp)=\partial\ka_\pp/\partial\eps(0,\pp)$ makes sense, since $E$ is a decreasing bijection: if $\eps_\pp$ is the unique value for which $E(\eps_\pp)=\pp$ then $[0,\eps_\pp]\subset \kasetc$.}
%Hence,
%$\ka_{\pp\eps}(\eta,\pp)>0,\,\forall\eta\in[0,\epss),\,\forall\pp\in[0,E(\eps)]$.
By $\partial\ka^*/\partial\pp(0,\pp)=\ka'_0(\pp)=1$ we obtain, for any $\eps\in(0,\epss]$ and $\pp\in[0,E(\eps)]$,
\be
\ka'_\eps(\pp)-1=\fpd{\ka^*}{\pp}(\eps,\pp)-\fpd{\ka^*}{\pp}(0,\pp)=\int_0^\eps K_{\pp\eps}(\eta,\pp)d\eta>0.%\quad \forall \pp\in (0,E(\eps)),
\ee
Thus there exists a maximal number $\epsp\geq \epss>0$ for which $\ka_\eps$ is LIF-like, for any $\eps\in(0,\epsp]$. In the neural case we already remarked that \eqref{LIF1} implies \eqref{LIF-2} for all $\eps\in(0,\epsm]$ (trivially including the limit case $\eps=\epsm$).
\end{proof}

\begin{rem}\label{remark: LIF vs LIF-2} Taking the partial derivative of \eqref{Z-by-z} to $\pp$ and considering the limit $\eps\ra 0$, it follows that if $\ka_\eps$ is LIF-like for $\eps\in(0,\epsp]$
then $Z'(\pp)\geq 0$ for all $\pp\in[0,1]$. This is a weak converse of Proposition \ref{prop: LIF-like}.
%Using the proof technique of Theorem \ref{prop: main} below (replacing $J(\eps,\phi)$ by $\partial\kappa/\partial\pp(\eps,\pp)$), one can show that if a PTC $\ka$ is LIF-like then so are all corresponding PTC$_\eps$'s for sufficiently small $\eps$.
\end{rem}

We emphasize that we do {\em not} adopt the assumption of LIF-like PTC$_\eps$'s or iPRC's in the present paper.
%However, we will compare our results with those obtained in \cite{MirStr90} by using the characterization of LIF-like (neural) models as defined above.

\subsection{Phase space and firing map}\label{subsec: phase space}

Our approach to a better understanding of the network dynamics is based on the
%of an IF network $(N,f,\eps)$
{\em firing map}, as introduced by Mirollo and Strogatz~\cite{MirStr90}, and briefly recalled in this section.  We assume a fully connected network of oscillators with identical and non-additive (not necessarily neural) coupling and formulate a precise setting in the case of excitatory coupling.
% the setting for inhibitory coupling being obtained by evident replacements. %Differently from
Unlike \cite{MirStr90}, we sample the cluster phases
right {\em before} (instead of {\em after}) the firing events.

Suppose that the network consists of $l+1$ oscillator clusters right before the next firing. If $l=0$ there is only one cluster: the system remains synchronized because of all-to-all coupling.%and resulting in $h_\eps_0=0$, where $h_\eps$ is the firing map fully defined in the sequel.

Assume now $l\geq 1$ and define
\begin{equation}\label{def-Sl}
S_l\equiv \{\th=(\th_1,\ldots,\th_l)\in \R^l|0<\th_1<\ldots<\th_l<1\}.
\end{equation}
The cluster that is about to fire has phase  $\th_{l+1}=1$, and the system is characterized by the {\em reduced phase vector} $\th\in S_l$ of the phases, {\em ranked in ascending order}, of the remaining clusters. The cluster with the smallest phase at that instant has phase $\th_1$, the cluster with the second smallest phase has phase $\th_2$, and so on.

The firing of the cluster changes the phases of the other clusters, according to the {\em jump map} $\tau_{\eps,l}:S_l\ra\overline{S_l}$ defined by
\begin{eqnarray}\label{def-taul}
\tau_{\eps,l}(\th_1,\ldots,\th_l)=(\ka_\eps(\th_1),\ldots,\ka_\eps(\th_l)).
\end{eqnarray}
We partition the space $S_l$ according to the number of absorptions caused by this firing: if $l-k$ absorptions take place, $0\leq k\leq l$, then $\th$ belongs to $S_{\eps,lk}\subset S_l$:
\begin{eqnarray}\label{def-Slk}
S_{\eps,lk}&\equiv&\{\th\in S_l|k\;\textrm{is the largest index $i$ s.t.}\;\ka_\eps(\th_{i})<1\}\\
&=&\tau_{\eps,l}^{-1}(A_{lk}),\\
A_{lk}&\equiv&\{\th\in \overline{S_l}|0<\th_1<\ldots<\th_k<\th_{k+1}=\ldots=\th_{l}=1\}.
%=1\equiv\ka_\eps(\th_{l+1})
\end{eqnarray}
%where $A_{ll}=S_l$.
Notice that $S_{\eps,ll}$ is open since $A_{ll}=S_l$ is open and $\tau_{\eps,l}$ is continuous. Right after the firing, the phase of the
%oscillator
cluster consisting of the oscillators that have fired or have been absorbed is reset from 1 to 0. If $\th\in S_{\eps,lk}$, the  reduced phase vector of the $k$ non-absorbed clusters is $\tau_{\eps,lk}(\th)$, where
\begin{eqnarray}\label{def-taukl}
\tau_{\eps,lk}\equiv\pi_{lk}\circ\tau_{\eps,l}|_{S_{\eps,lk}}:\quad S_{\eps,lk}\ra S_k;\quad (\th_1,\ldots,\th_l)\mapsto (\ka_\eps(\th_1),\ldots,\ka_\eps(\th_k)),
\end{eqnarray}
with $\pi_{lk}:\R^l\ra\R^k$  the standard projection map $(\th_1,\ldots,\th_l)\mapsto (\th_1,\ldots,\th_k)$. Recall that $\k_\eps$ is strictly increasing; this implies that the components $\ka_\eps(\th_1),\ldots,\ka_\eps(\th_k)$ are still in ascending order. In other words  $\tau_{\eps,lk}(\th)$ belongs to $S_k$, and in the case $k=l$ where no absorptions take place, $\tau_{\eps,ll}:S_{\eps,ll}\ra \tau_{\eps,ll}(S_{\eps,ll})$ is a diffeomorphism between open sets. %, with inverse $\tau_{-\eps,kk}$.
%\begin{eqnarray}\label{def-taul}
%\tau_{\eps,ll}:\quad S_{l}\rightarrow S_l;\quad (\th_1,\ldots,\th_l)\mapsto (\ka_{-\eps}(\th_1),\ldots,\ka_{-\eps}(\th_l)).
%\end{eqnarray}
We formally introduce the space $S_0=S_{\eps,00}=\R^0=\{0\}$ and define $\tau_{\eps,00}(0)= 0$.
Notice that for any $0\leq k\leq l$ one can rewrite $\tau_{\eps,lk}$ as
\begin{equation}\label{write-taukl}
\tau_{\eps,lk}=\tau_{\eps,kk}\circ \pi_{lk}.
\end{equation}

When $k=0$ then the system has synchronized. If $k\geq 1$, the phases of the $k+1$ clusters evolve according to the dynamics \eqref{phasedyn} until right before the next firing (by the cluster corresponding to the largest non-trivial phase $\k_\eps(\th_k)$ just after the last firing.). At this instant, the phase vector of the remaining clusters has changed to
%\be\label{def-heps-kl}
%h_{\eps,kl}(\pph)\equiv
$\s_k(\tau_{\eps,lk}(\th))$,
%\ee
where $\s_k$ is the affine {\em shift map}
\begin{equation}\label{def-si-l}
\s_k:\quad \R^k\rightarrow \R^k;\quad (\th_1,\ldots,\th_k)\mapsto (1-\th_k,\th_1+1-\th_k,\ldots,\th_{k-1}+1-\th_k),
\end{equation}
%with inverse
%\begin{equation}\label{def-si-l-inv}
%\s_l^{-1}:\quad \R^l\rightarrow \R^l;\quad (\th_1,\ldots,\th_l)\mapsto (\th_2-\th_1,\th_3-\th_1,\ldots,1-\th_1).
%\end{equation}
and defining $\s_0(0)=0$. Notice that the restriction of $\s_k$ to $S_k$ is an auto-diffeomorphism.

Let $n\equiv N-1$ for a network  of $N$ initial clusters.
%For a network $(N,f,\eps)$ the maximal value for $k$ is $n\equiv N-1$.
We introduce a global jump map $\tau_\eps$ and a global shift map $\s$, defined on the {\em phase space}
\be\label{phase space}
S=\bigcup_{l=0}^n S_l=\bigcup_{l=0}^n\bigcup_{k=0}^l S_{\eps,lk}
\ee
and acting as $\tau_{\eps,lk}$ and $\s_l$ on $S_{\eps,lk}$ and $S_l$, respectively.
The {\em firing map} of the network, transforming the phase vector right before a firing to the reduced phase vector right before the next firing, is thus given by
\begin{equation}\label{def-heps}
h_\eps:\quad S\rightarrow S;\quad \th\mapsto \s(\tau_\eps(\th)).
\end{equation}
This defines a discrete-time dynamical system on $S$, with $S_n$ as the space of {\em initial reduced phase vectors}  or {\em initial conditions} which will be denoted as $\ph=(\ph_{1},\ldots,\ph_{n})$. Notice that $h_\eps$, when restricted to $S_{\eps,nn}$, acts as a diffeomorphism between open sets:
\be\label{heps-nn}
h_{\eps,nn}\equiv\s_n\circ\tau_{\eps,nn}:\quad S_{\eps,nn}\ra h_{\eps,nn}(S_{\eps,nn}).
\ee

%~\footnote{The dynamical system is non-classical in the sense that vectors at different times may belong to different $\R^k$. This could be circumvented by working with the union $S^*\subset\R^n$ of the spaces $S^*_k\equiv \{\th\in\R^n|\th_1<\ldots<\th_k<\th_{k+1}=\ldots=\th_n=1\equiv \th_{N+1}\},\,0\leq k\leq n$ instead and defining maps $\tau:(\th_i)\mapsto (\ka_\eps(\th_i))$ and $\s:\th\in S^*_k\mapsto (1-\th_k,\th_1+1-\th_k,\ldots,\th_{k-1}+1-\th_k,1,\ldots,1)$ on $S^*$. However, this would make some definitions in the sequel more demanding.}

%\begin{equation}
%h^{i}:\quad \R_{\geq 0}\times S\rightarrow S;\quad (\eps,\th)\mapsto (\eps,h_\eps^i(\th)),\quad h_\eps^i(\th)\equiv %(\underbrace{h_\eps\circ\ldots\circ h_\eps}_{i})(\th).
%\end{equation}

%For later use we define the map %for all $i\in\N$ the maps
%\begin{equation}
%h:\quad \X\times S\rightarrow [-\epsm,\epsm]\times S;\quad (\eps,\th)\mapsto (\eps,h_\eps(\th)).
%\end{equation}

%notice: if the coupling is not all to all and the oscillators are merely labelled from $1$ to $N$, then this is no longer true, since $\theta^{**}_k$ is no longer the largest value now.

\section{Main theorems}\label{sec: main}

We focus on synchronization of a network of all-to-all and identically coupled oscillators.
%Consider an initial condition $\ph=(\ph_{1},\dots,\ph_{n})\in S_{n},\;n=N-1$ corresponding to  the initial phases of the oscillators  and listed in ascending order. Without loss of generality we assume that $\ph_{N}=1$, corresponding to the $N$-th oscillator ready to fire.
Suppose that there are $N$ initial oscillator clusters.
% of oscillators with identical phases.
%Due to the identical unperturbed dynamics,
We may assume that one of these clusters has phase $1$ and is ready to fire. Consider the corresponding initial condition $\ph=(\ph_{1},\dots,\ph_{n})\in S_{n},\;n=N-1$, being the vector of phases of the other oscillator clusters listed in ascending order.
As explained in the previous section the dynamics of this network consists of phase advances (modeled by $\tau_{\eps}$) triggered by a firing cluster, alternated with all oscillators moving uniformly until the next firing (modeled by $\s$).  The evolution in time is then nicely captured by repeated application of the firing map $h_\eps=\s\circ\tau_{\eps}$ starting in $\phi$.

We set conditions on the iPRC of the underlying oscillators such that synchronization
%(i.e. all phases equal)
is brought about for large enough time, for almost all initial conditions. Synchronization is a consequence of a  process of continuing absorption into increasing clusters of oscillators with identical phases, as in  \cite{MirStr90}. We introduce a comprehensive set of conditions  for absorption that are weaker in general, and agree with the conditions suggested  for the specific
%network of oscillators
LIF-like case studied in \cite{MirStr90}.

Absorption plays a crucial role in our contribution and will be studied first. The synchronization  result  follows using arguments  along the lines of  \cite{MirStr90}.

We always consider the standard Euclidean topology on $\R^k$, where $\overline{X}$ denotes the closure of a set $X$. The map $f^i$ denotes the
$i{\textrm{-fold}}$ composition of a map $f:X\ra X$, and $f^{-i}(X)$ denotes the inverse image of $X$ under  $f^i$.
% wrt this topology.
%, and $h^{-i}(X)$ symbolizes the inverse image of $X$ under the $i^{\textrm{th}}$ power of the map $h$.
The symbol $\mu_k$ stands for the Lebesgue measure on $\R^k$.

\subsection{Absorption}\label{subsec: absorption}

For any fixed $\eps\in [0,\epsm]$ and integer $i\geq 0$ we define the set of initial conditions $\varphi \in S_{n}$ that {\em survive} $i$ applications of $h_\eps$ without any absorption:~\footnote{This set was denoted by $A_{i+1}$ in \cite{MirStr90}, but the notation $A_{\eps,i}$ is more convenient for our purposes.}
\begin{equation}\label{def-Ar}
A_{\eps,i}\equiv \{\ph\in S_n|h^j_{\eps}(\ph)\in S_n,\forall j=0,\ldots, i\}.
\end{equation}
%Notice that $A_{0,i}=S_n$ for any $i\geq 0$, $A_{\epsm,0}=S_n$ and $A_{\epsm,i}=\emptyset$ for any $i\geq 1$, and
It follows from the definition that
\be\label{conseq-Ai}
A_{\eps,0}=S_n,\qquad A_{\eps,1}=S_{nn},\qquad \forall\; i\geq 1:\;A_{\eps,i}\subset A_{\eps,i-1}.%\quad h_\eps(A_{\eps,i})\subset A_{\eps,i-1}.
\ee
Since the firing map $h_\eps$ acts as $h_{\eps,nn}$ as long as no absorptions take place, and since $h^i_{\eps,nn}(\ph)\in S_n$ implies $h^j_{\eps,nn}(\ph)\in S_n,\forall j\leq i$, we obtain for all $i\geq 0$:
\begin{equation}\label{prop-Ar}
A_{\eps,i}=\{\ph\in S_n|h^i_{\eps,nn}(\ph)\in S_n\}=h_{\eps,nn}^{-i}(S_n).
\end{equation}
This implies $A_{\eps,i}=h_{\eps,nn}^{-1}(A_{\eps,i-1})$, whence  $h_{\eps,nn}(A_{\eps,i})\subset A_{\eps,i-1}$ and $h_{\eps,nn}^j(A_{\eps,i})\subset A_{\eps,i-j}$ for any $j\leq i$.
Also, $A_{\eps,i}\subset S_n$ is open since $S_n$ is open and $h_{\eps,nn}$ is continuous. As a generalization of \eqref{heps-nn}, we conclude that for any $j\leq i$ the restriction $h_\eps^j|_{A_{\eps,i}}$ acts as a diffeomorphism $h_{\eps,nn}^j$ between the open sets $A_{\eps,i}$ and $h_{\eps,nn}^j(A_{\eps,i})$.
Taking $j=N$, we define the {\em return map} as
\begin{equation}\label{def-Reps}
R_\eps\;\equiv\; h_{\eps,nn}^N:A_{\eps,N}\ra S_n,
\end{equation}
which thus acts as a diffeomorphism between the open sets $A_{\eps,i}$ and $R_\eps(A_{\eps,i})\subset A_{\eps,i-N}$  for all $i \geq N$.
When the pulse coupling is switched off ($\eps = 0$), then $\tau_{0,nn}=\id$ and therefore
$R_0=\id$ since $\s_n^N=\id$.

The set of initial conditions for which no
%single
absorption takes place as the system evolves in time is given by:
\begin{equation}\label{def-Aeps}
A_\eps\equiv \bigcap_{i=0}^\infty A_{\eps,i}.%=\lim_{i\ra\infty} .
\end{equation}
We are now ready to prepare the ground for the main result on absorption. Our goal is to develop reasonable conditions on the model parameters  such that $A_\eps$ is very small, more precisely such that $\mu_n(A_{\eps})=0$.  %with $\mu_n$  the Lebesgue measure on $\R^n$.
First we prove a fundamental result on the evolution of the volume of $A_{\eps,i}$ under application of $R_{\eps}$ in terms of its Jacobian determinant, which  is explicitly computed in terms of its composing functions. In a second step we make a crucial observation.

\subsubsection{Jacobian determinant of $R_{\eps}$}\label{subsubsection: jacobian}

\begin{prop}\label{lem: 1-Jac}%~\footnote{See \cite{WyllAey14} for a more general version of this lemma.}
Suppose that the Jacobian determinant
of $R_\eps$ satisfies
\begin{equation}\label{Jac-cond}
\det(D R_\eps)(\ph)\geq 1+a_\eps,\quad a_\eps\in\R_{>0},\qquad \forall \ph\in A_{\eps,N}.
\end{equation}
Then $\mu_n(A_\eps)=0$.
\end{prop}

\begin{proof} As explained above $R_\eps(A_{\eps,kN})\subset A_{\eps,(k-1)N}$, and  taking $j=N$ and $i=kN$,   $R_\eps$ acts as a diffeomorphism between the open sets $A_{\eps,kN}$ and $R_\eps(A_{\eps,kN})$, for any $k\geq 1$. From which we obtain
\begin{eqnarray}
\mu_n(A_{\eps,(k-1)N})
&\geq& \mu_n(R_\eps(A_{\eps,kN}))=\int_{A_{\eps,kN}}|\text{det}(D R_\eps)(\ph)|\d\ph\label{qq}\\
&\geq&(1+a_{\eps})\int_{A_{\eps,kN}}\d\ph=(1+a_\eps)\mu_n(A_{\eps,kN}),\nonumber
\end{eqnarray}
where $\d\ph\equiv \d\ph_1\cdots\d\ph_n$; The last inequality follows from the assumption
%on $A_{\eps,N}$
\eqref{Jac-cond}
and since $A_{\eps,kN}\subset A_{\eps,N}$. By induction,
\be\label{ineq mu-n}
\mu_n(A_{\eps,kN})\leq \frac{\mu_n(A_{\eps,(k-1)N})}{1+a_\eps}\leq \frac{\mu_n(A_{\eps,0})}{(1+a_\eps)^k}=\frac{1}{n!(1+a_\eps)^k},
\ee
since the volume of $A_{\eps,0}=S_n$ is $1/n!$. Hence,
\be
\mu_n(A_\eps)=\lim_{i\ra\infty}\mu_n(A_{\eps,i})=\lim_{k\ra\infty}\mu_n(A_{\eps,kN})=0,
\ee
where the first equality is due to a well-known property of the measure  of a descending chain of sets $(A_{\eps,i})_{i=0}^\infty$, and the last equality follows from \eqref{ineq mu-n}.
\end{proof}
We denote $\det(D R_\eps)(\ph)$ by $J(\eps,\ph)$ and write an explicit expression. For any $\ph\in A_{\eps,N}$,
\beq
J(\eps,\ph)
&=&\prod_{i=0}^n \det(D h_{\eps,nn}(h_{\eps,nn}^i(\ph)))
=\prod_{i=0}^n \det(D\s_n(\tau_{\eps,nn}(h_{\eps,nn}^i(\ph))))\det(D \tau_{\eps,nn}(h_{\eps,nn}^i(\ph)))\nonumber\\
&=&\prod_{i=0}^n\det(M)\prod_{j=1}^n \ka'_\eps(h_{\eps,nn}^i(\ph)_j)\label{pr}\\
&=&\prod_{i=0}^n\prod_{j=1}^n \fpd{\k}{\pp}(\eps,h_{ij}(\eps,\ph))\label{tr}
\eeq
The first and second equalities follow from the chain rule. $M$ denotes the constant Jacobian matrix of the affine map $\s_n$, and the definition \eqref{def-taukl} of $\tau_{\eps,nn}$ is invoked to obtain the third equality. The last equality is based on the fact that $M^{n+1}$ is the identity matrix in $n$ dimensions (since $\s_n^{N}$ is the identity map on $\R^n$), definition \eqref{def-k}, and
\be\label{def-hij}
h_{ij}(\eps,\ph)\equiv h^i_{\eps,nn}(\ph)_j,
\ee
where $i$ denotes the number of compositions and $j$ refers to the $j$-th component.

\begin{cor}\label{cor: LIF-like} Suppose that a network of pulse-coupled oscillators has an LIF-like PTC$_\eps$. Then $\mu_n(A_\eps)=0$.
\end{cor}

\begin{proof}
%The continuous extension of the function $\ka_\eps'(.)-1$ to the compact set $[0,E(\eps)]$  reaches a minimum value, $b_\eps>0$, and by \eqref{tr} the condition \eqref{Jac-cond} is fulfilled by $a_\eps=(1+b_\eps)^{Nn}-1$.
Let $\ka_\eps$ denote the PTC$_\eps$. By definition \ref{definition: LIF}, the function $\ka_\eps'(.)-1$ takes a minimum value $b_\eps>0$ on the compact set $[0,E(\eps)]$. Suppose now that $\ph\in A_{\eps,N}$. Then $\ph$ survives $N$ applications of $h_\eps$ without any absorption, such that the arguments $h_\eps^i(\ph)_j$ of $\ka_\eps'$ in \eqref{pr} all belong to $(0,E(\eps))$. It follows from \eqref{pr} and $\det(M)=1$ that \eqref{Jac-cond} is fulfilled by $a_\eps=(1+b_\eps)^{Nn}-1$.
\end{proof}

%\begin{proof}
%Suppose that $\ph\in A_{\eps,N}$. Then $\ph\in A_{\eps,2}$, whence
%\be\label{A2-cond}
%h_{\eps}(\ph)_n=\ka_\eps(\ph_{n-1})+1-\ka_\eps(\ph_n) <E(\eps).
%\ee
%On the one hand, this implies $\ka_\eps(\ph_n)> 1-E(\eps)+\ka_\eps(\ph_{n-1})> 1-E(\eps)$. By assumption, $k_\eps'>1$ on $(0,E(\eps))$, such that $k_\eps(0)>1-E(\eps)$ by $\ka_\eps(E(\eps))=1$ and the Mean Value Theorem; since $E(\eps)>0$ and $\ka_\eps$ strictly increases on $(0,E(\eps))$, the Intermediate Value Theorem implies the existence of a unique phase value $\pp_1\in (0,1)$ for which $\ka_\eps(\pp_1)=1-E(\eps)$. Hence, $\ph_n> \pp_1$.
%On the other hand, \eqref{A2-cond} implies $\ka_\eps(\ph_{n-1})< E(\eps)+\ka_\eps(\ph_{n})-1< E(\eps)$. If $E(\eps)\leq\ka_\eps(0)$ then $E(\eps)<\ka_{\eps}(\pp)$ for all $\pp\in(0,1)$ since $\ka$ increases, and $A_{\eps,2}=A_{\eps}=\emptyset$. If $E(\eps)>\ka_\eps(0)$ then the Intermediate Value Theorem implies  the existence of a unique phase value $\pp_0\in(0,1)$ for which $\ka_\eps(\pp_0)=E(\eps)$. Hence, $\ph_{n-1}<\pp_0$.
%If $E(\eps)> 1-E(\eps)$ then $\ph_1<\ph_{n-1}<\ph_n<\ph_0$. The function $\ka_\eps'(.)-1$ takes a minimum value $b_\eps>0$ on the compact set $[\ph_1,\ph_0]$. We have $J(\eps,\ph)\geq k'_\eps(\ph_n)k'_\eps(\ph_{n-1})\geq (1+b_\eps)^2$, for any $\ph\in A_{\eps,N}$, and \eqref{Jac-cond} is fulfilled for $a_\eps=b_\eps(b_\eps+2)>0$.
%If $E(\eps)\leq 1-E(\eps)$ then we can't conclude anything... . Dus we gaan ons er niet meer mee bezighouden.
%\end{proof}

\begin{rem} Corollary \ref{cor: LIF-like} formulates Theorem 3.1 by Mirollo and Strogatz~\cite{MirStr90} for an LIF-like network with PTC$_\eps$ not necessarily of the neural form \eqref{def-keps}. As in \cite{MirStr90}, the corollary holds for any positive pulse strength $\eps$.
\end{rem}

\begin{rem}
 %such that the parameter $\eps$ is superfluous.
In the proof of Theorem 3.1 of \cite{MirStr90}, the inclusion $R_\eps(A_\eps)\subset A_\eps$ is invoked.
%, with $A=A_\eps$ and $R=R_\eps$.
The argument in \cite{MirStr90} is based on a calculation of the Jacobian of $R_\eps$ {\em within} $A_\eps$ (see also pp.\ 27-29 of \cite{Strogatzbook}), and invoking the equality in \eqref{qq}. However, this equality holds and makes sense only when $R_\eps$ acts on an open subset, while $A_\eps$ is not open in general. Hence, the proof of Theorem 3.1 of \cite{MirStr90} leads to the conclusion that {\em $A_\eps$ does not contain any non-empty open subset}.~\footnote{Suppose that the largest open subset $K$ of $A_\eps$ is non-empty. Then $R_\eps(K)\subset K$, implying $\mu_n(R_\eps(K))\leq \mu_n(K)$. However, $\mu_n(R_\eps(K))=\int_{K}|\text{det}(D R_\eps)(\ph)|\d\ph\geq (1+a_\eps)\mu_n(K)>\mu_n(K)$, a contradiction.} This is a weaker statement than $\mu_n(A_\eps)=0$;  examples like the fat Cantor set illustrate this clearly as they have non-zero measure but empty interior. However, by looking at $A_{\eps}$ as the intersection of a descending chain of {\em open} sets $A_{\eps,i}$, see
\eqref{conseq-Ai} and
\eqref{def-Aeps}, and using the proof technique of Proposition \ref{lem: 1-Jac}, we recovered $\mu_n(A_\eps)=0$.
%as illustrated in the case $n=1$ by examples like the fat Cantor set ${\cal C}_f$, which does not contain any open subsets but has $\mu_1({\cal C}_f)=1/2$.
\end{rem}

The proof of the main theorem below is based on securing \eqref{Jac-cond} for sufficiently small $\eps$, by exploring the limit case $\eps=0$. Referring to \eqref{pr}, this means evaluating scalar functions in the points
\be\label{h0i}
h_0^i(\ph)_j\,=\,h_{ij}(0,\ph)\,=\,\s^i_n(\ph)_j,\quad i=0,\dots, n,\,j=1,\dots,n.
\ee
We first make a crucial observation on the multiset consisting of these points.
%, where $\ph$ belongs to the closure $\overline{S_n}$ of $S_n$.

%{\bf We shall consider a collection of networks modeled by a PTC $\ka\equiv \{\ka_\eps\}$. By Proposition \ref{prop: LIF-like} and Corollary \ref{cor: LIF-like}, absorption for almost all initial conditions and for weak pulse coupling is guaranteed by the LIF-like condition \eqref{intro:mirollostrogatz}. We will show that it is in fact guaranteed by the weaker condition \eqref{intro:maincondition}.}

%\subsubsection{Explicit computation}\label{subsubsection: computation}

\subsubsection{Crucial observation}\label{subsubsection: observation}

We represent the $N$ initial oscillator clusters by points on a circle. Without loss of generality, one cluster is in the origin, receiving  label  $0$ and assigned phase $\phi_0=0$. Let $\phi=(\phi_1,\ldots,\phi_n)\in S_n$ be the initial reduced phase vector, where $0<\phi_1<\ldots<\phi_n<1$. Consider now the limit of zero pulse strength, $\epsilon=0$. In this case, the dynamics amounts to cluster points merely shifting at each iteration, and after $i$ iterations the reduced phase vector has become $h_0^i(\phi)=\sigma_n^i(\phi)$. Letting $i$ run from $0$ to $n$, each cluster `fires' exactly once, and we obtain $Nn$ components $\sigma^i_n(\phi)_j,\, i=0,\dots, n,\,j=1,\dots,n$. In an equivalent, `passive' picture one leaves the points representing the initial clusters fixed and considers, for each cluster point, the arcs measured counterclockwise from that point to the $n$ remaining points; the $Nn$ components $\sigma^i_n(\phi)_j,\, i=0,\dots, n,\,j=1,\dots,n$ are then recovered as the normalized lengths of these arcs. In this picture we notice that, for $k<l$, the arc length is $\phi_l-\phi_k$ when $k$ `fires', and $1-(\phi_{l}-\phi_{k})$ when $l$ `fires'.

%This observation also holds if some of the components $\ph$ are equal or have value $0$ or $1$. In other words, it is valid for any $\ph$ in the closure $\overline{S_n}=\{\ph\in\R^n|0\leq \ph_1\leq \ldots\leq \ph_n\leq 1\}$ of $S_n$.

\begin{figure}[h!]
    \begin{center}
        \subfigure{
            \includegraphics[width=0.4\linewidth]{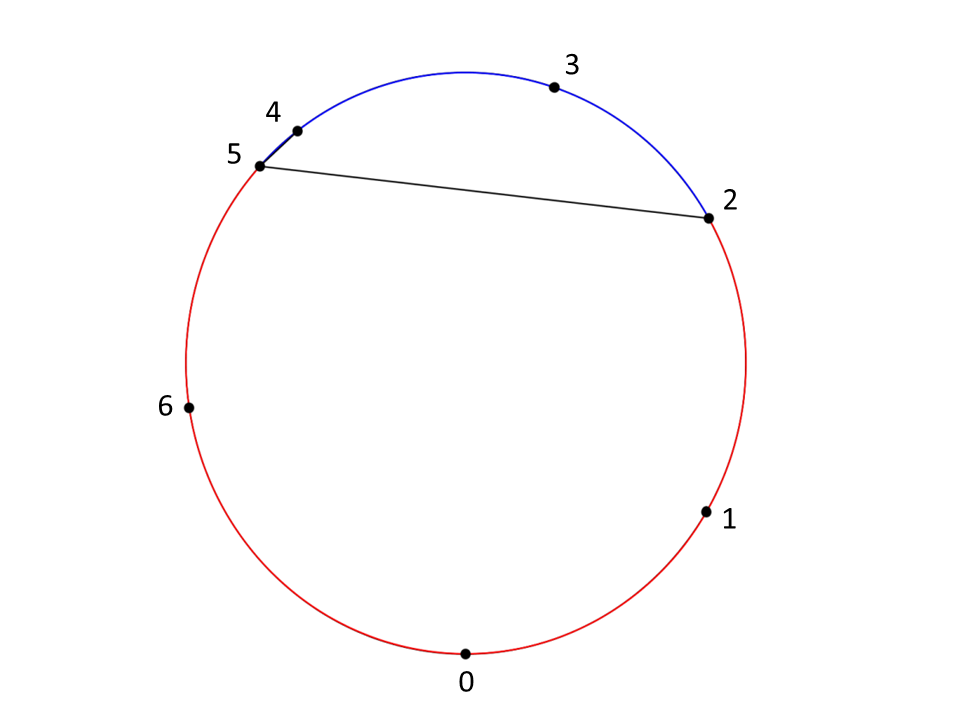}
        }
    \end{center}
    \caption{Representation on the circle of an example with 7 initial oscillator clusters. In case the pulse strength is $0$ and in the passive picture, the cluster points don't move. When cluster 2 `fires', cluster 5 has a relative phase $\ph_5-\ph_2$ (length of the blue arc), while the relative phase of cluster 2 is $1-(\ph_5-\ph_2)$ when cluster 5 `fires'   (length of the red arc). Similarly for all other connections.}
    \label{fig: observation}
\end{figure}

In conclusion, the  multiset of the $Nn$ components $\sigma^i_n(\phi)_j,\, i=0,\dots, n,\,j=1,\dots,n$ is the multiset sum of $Nn/2$ pairs, formed by the normalized arc length $\phi_{l}-\phi_{k}$ between two
%(possibly equal, or $0$-valued or $1$-valued)
initial phases and its complement $1-(\phi_{l}-\phi_{k})$ with respect to the normalized length of the circle:
\be\label{set-sigmas}
\forall \ph\in S_n:\quad \{\sigma^i_n(\ph)_j|i=0,\dots, n,\,j=1,\dots,n\}=\biguplus_{0\leq k<l\leq n}\{\phi_{l}-\phi_{k},1-(\phi_{l}-\phi_{k})\},
\ee
where $\ph=(\ph_1,\ldots,\ph_n)$ and $\ph_0=0$. For instance, if $N=4$ and $\phi=(1/4,1/2,3/5)$ then this multiset consists of 12 elements, viz., single elements $3/5,2/5,7/20,13/20,1/10,9/10$ and double elements $1/4,3/4,1/2$. In Figure \ref{fig: observation} an example with 7 initial oscillator clusters is depicted.
%For instance, if $N=5$ and $\ph=(0,a,b,b)$ then this multiset consists of 20 elements, viz., double elements $0,1,a,1-a,b-a,1-(b-a)$ and quadruple elements $b,1-b$.

%In Figure \ref{fig: observation} an example of 7 initial oscillator clusters

\subsubsection{Main result}\label{subsection: main result absorption}

Consider the extreme cases $\eps=0$ and $\eps=\epsm$ in definitions \eqref{def-Ar} and \eqref{def-Aeps}: on the one hand we have
%$A_{0,i}=S_n$ for any $i\geq 0$, whence $A_0=S_n$ and $\mu_n(A_0)=1/n!$.
$$
A_{0,i}=S_n,\quad \forall i\geq 0, \qquad A_0=S_n,\qquad \mu_n(A_0)=1/n!,
$$
while on the other hand,
%$A_{\epsm,i}=\emptyset$ for any $i\geq 1$, such that $A_{\epsm}=\emptyset$ and $\mu_n(A_{\epsm})=0$.
$$
A_{\epsm,i}=\emptyset,\quad \forall i\geq 1, \qquad A_{\epsm}=\emptyset,\qquad \mu_n(A_{\epsm})=0.
$$
We now show that the condition \eqref{intro:maincondition} guarantees $\mu_n(A_\eps)=0$ for sufficiently small pulse strengths $\eps\neq 0$.
%in some interval $(0,\epsp]$, with $\epsp>0$.
The argument is close to the proof of
%is similar to the one for
Proposition \ref{prop: LIF-like}.

%We start with the following:

%\begin{lem}\label{lem: phi-eps} For any given $\ph\in S_n$ and $i\in\N$, there exists a positive real number $\eps_{\ph,i}\leq \epsm$ such that $\ph\in A_{\eps,i}$, for all $\eps<\eps_{\ph,i}$.
%\end{lem}

%\begin{proof} By induction on $i$. Since $A_{\eps,0}=S_n$ for all $\eps\in[0,\epsm]$ the statement is trivial for $i=0$, with $\eps_{\ph,0}=\epsm$. Suppose that the statement is true for $i-1$. Then $\ph\in A_{\eps,i-1}$ for any $\eps<\eps_{\ph,i-1}$, such that $\psi=h_\eps^{i-1}(\ph)$ belongs to $S_n$, implying $0<\psi_n<1$. Since the map $E$ is a decreasing bijection from $(0,\epsm)$ to $(0,1)$, a unique value $\eps_s$ exists such that $E(\eps)>E(\eps_s)=\psi_n$ for all $\eps<\eps_s$. Thus $\ka_\eps(\psi_n)<1$ and $\ph$ also belongs to $A_{\eps,i}$ for all $\eps<\eps_{\ph,i}\equiv \min(\eps_{\ph,i-1},\eps_s)$.
%\end{proof}

%For arbitrary $i\in\N$ this lemma implies that $A_{\eps,i}$ is non-empty if $\eps$ is smaller than~\footnote{If there did not exist $\ph\in A_{\eps,i}$ if $\eps<\eps_i$ then, by Lemma \ref{lem: phi-eps}, $\eps\geq \eps_{\ph,i}$ for all $\ph\in S_n$, whence $\eps\geq \eps_i$ by applying the supremum function to this inequality, a contradiction.}
%\be
%\eps_i\equiv \sup\{\eps_{\ph,i}|\ph\in S_n\}.
%\ee

%Opmerking: zolang $\eps$ voldoende klein is, is $A_{\eps,N}\neq \emptyset$. Inderdaad, in dit geval bezit $h_\eps$ een evenwichtspunt, wat automatisch een element is van $A_{\eps,\infty}\subset A_{\eps,N}$.

\begin{thm}\label{prop: main}
Suppose that the PTC $\ka$
%and its partial derivatives satisfy the continuity and extendability requirements described in Sec.\ \ref{subsec: oscillator}.
satisfies properties P1-P4 outlined in  Sec.\ \ref{subsec: oscillator}.
If the corresponding iPRC\, $Z$ satisfies
\be\label{Z-cond-p}
Z'(\pp)+Z'(1-\pp)>0,\qquad \forall \pp\in[0,1],
\ee
then there exists $\epsp>0$ such that $\mu_n(A_\eps)=0$ for any $\eps\in(0,\epsp]$.
\end{thm}

\begin{proof} Let $\eps\neq 0$ be an arbitrary pulse strength for which $A_{\eps,N}\neq\emptyset$. The map $\det(D R_\eps)-1=J(\eps,.)-1$ has domain $A_{\eps,N}$. By \eqref{tr} and the continuous extendability of $\ka$ and $\partial{\ka}/\partial{\pp}$ (property P4 in Sec.\ \ref{subsec: oscillator}) there exists a unique continuous extension of $J(\eps,.)-1$ to $\overline{A_{\eps,N}}$, which we still denote by the same symbol. Since $\overline{A_{\eps,N}}$ is compact, this continuous extension takes a minimum value $a_\eps$. If $J(\eps,.)-1$ is {\em strictly positive} on $\overline{A_{\eps,N}}$ then $a_\eps>0$, and we find $\mu_n(A_\eps)=0$ by applying Proposition \ref{lem: 1-Jac}. Hence, there remains to be shown that
\be\label{whatweneed}
\exists \eta>0,\,\forall \eps\in(0,\eta]:\quad (A_{\eps,N}=\emptyset)\quad\textrm{or}\quad (\textrm{$A_{\eps,N}\neq\emptyset$ and $J(\eps,.)-1>0$ on $\overline{A_{\eps,N}}$}).
\ee
%an interval $(0,\eta]$ exists such that, for all pulse strengths $\eps$ belonging to this interval, either $A_\eps=\emptyset$, or $A_\eps\neq\emptyset$ and $J(\eps,.)-1$ is strictly positive on $\overline{A_{\eps,N}}$.
Then $\epsp$ in the statement of the theorem is taken to be the maximal such $\eta$.

%Hence, we need to show that the set of pulse strengths
%\be\label{ssset}
%\{\eta|\forall \eps\in (0,\eta]:\;(A_\eps=\emptyset)\;\textrm{or ($A_\eps\neq\emptyset$ and $J(\eps,.)-1$ is strictly positive on $\overline{A_{\eps,N}}$)}\}
%\ee
%is non-empty (meaning that the set contains an interval ). Then $\epsp$ in the statement of the theorem is taken to be the maximal element of this set.

In accordance with property P4, take any continuous extension $\ka^*$ of $\ka$ to $\R^2$, with continuous partial derivatives up to second order. For any $\eps\in\R$ define a  real map $\ka^*_\eps$ on $\R$ by $\ka^*_\eps(\pp)=\ka^*(\eps,\pp)$. This leads to real maps on $\R^n$ that are corresponding continuous extensions of $\tau_{\eps,nn}$ (or $\tau_{\eps,n}$), $h_{\eps,nn}$, and $R_\eps$, and to a real continuous extension $h^*_{ij}$ on $\R^{n+1}$ of $h_{ij}$, defined by (cf.\ \eqref{def-taul}, \eqref{def-taukl}, \eqref{heps-nn}, \eqref{def-Reps}, \eqref{def-hij}):
\bes
\tau^*_{\eps,nn}(\th)=(\ka^*_\eps(\th_1),\ldots,\ka^*_\eps(\th_n)),\qquad h^*_{\eps,nn}=\s_n\circ\tau^*_{\eps,nn},\qquad R^*_\eps= h^{*N}_{\eps,nn},\qquad h^*_{ij}(\th)=h^{*i}_{\eps,nn}(\th)_j.
\ees
The map $J^*:\R^{n+1}\ra\R$ defined by $J^*(\eps,\ph)=\det(D R^*_\eps)(\ph)$ is continuous and is equal to the expression \eqref{tr}  with $\ka$ and $h_{ij}$ replaced by their extensions. Defining
$$\F\equiv\fpd{J^*}{\eps}$$  %$\F\equiv\partial{J^*}/\partial{\eps}$
and invoking Schwarz's rule
we obtain
\bes%\label{dJdeps}
\F(\eps,\ph)=\sum_{i=0}^n\sum_{j=1}^n\left\{
\left[
\fpd{}{\pp}\left(\fpd{\ka^*}{\eps}\right)
(\eps,h^*_{ij}(\eps,\ph))+{\ka^*_\eps}''(h^*_{ij}(\eps,\ph))\fpd{h^*_{ij}}{\eps}(\eps,\ph)\right]\prod_{(k,l)\neq (i,j)}{\ka^*}'_\eps(h^*_{kl}(\eps,\ph))\right\}.
\ees
Take $\eps=0$ and $\ph=(\ph_1,\ldots,\ph_n)\in S_n$, and define $\ph_0=0$. By \eqref{Z-prop}, \eqref{h0i}, the crucial observation \eqref{set-sigmas}, and $\ka_0=\id$, the above expression reduces to
\bes
\F(0,\ph)=\sum_{0\leq k< l\leq n}[Z'(\ph_l-\ph_k)+Z'(1-(\ph_l-\ph_k))].
\ees
Notice that both sides of this equality define identical continuous functions on $S_n\ni\ph$ as well as on the closure $\overline{S_n}\ni\psi$, since $Z'$ is defined and continuous on $[0,1]$ and $\psi_l-\psi_k\in[0,1],\,0\leq k<l\leq n$ for $\psi\in \overline{S_n}$.
Moreover, the continuous function $\pp\mapsto Z'(\pp)+Z'(1-\pp)$ reaches a minimum, $a$, on the compact set $[0,1]$, with $a>0$ because of assumption \eqref{Z-cond-p}. Hence, for $\psi\in \overline{S_n}$ we obtain
\be\label{Jeps-epxr}
\F(0,\psi)=\sum_{0\leq k< l\leq n}[Z'(\psi_l-\psi_k)+Z'(1-(\psi_l-\psi_k))]\geq\frac{Nn}{2}a>0.%\geq \frac{Nnb}{2}.
\ee

The map $\F$ is continuous since obtained from continuous maps by continuity preserving operations. Hence the set $\F^{-1}((-\infty,0])$ is closed, and its intersection, $\Sp$, with the compact set $[0,\epsm]\times\overline{S_n}$ is compact. If $\Sp\neq\emptyset$
the continuous projection map
%$\textrm{pr}_\eps:
$(\eps,\psi)\mapsto\eps$ reaches a minimum $\epss$ on $\Sp$, with $\epss>0$ because of \eqref{Jeps-epxr}.
%which is strictly positive by \eqref{Jeps-epxr} (implying $\{0\}\times\overline{S_n}\nsubseteq\Sp$).
If $\Sp=\emptyset$ then set $\epss=\epsm$. It follows that $\F$ is strictly positive on $[0,\epss)\times\,\overline{S_n}$. Hence, whenever $A_{\eps,N}\neq\emptyset$ for $\eps\in (0,\epss]$, it follows for any $\psi\in \overline{A_{\eps,N}}\subset\overline{S_n}$ that
\be
J(\eps,\psi)-1=J^*(\eps,\psi)-J^*(0,\psi)=\int_0^\eps \F(s,\psi)ds>0,
\ee
since $J^*(0,\psi)=1$. This implies that $\eta=\epss$ satisfies \eqref{whatweneed}, as we needed to show. %, and $\epsp\geq\epss$.
%we found an interval $(0,\eta=\epss]$ as required. is an interval as required propertbelongs to the set \eqref{ssset} contains the which is thus non-empty, as we needed to show.
\end{proof}

\begin{rem} Whereas $\eps^*$ in the above proof depends on the chosen extension $\ka^*$, the value $\epsp$ and all values $a_\eps,\,\eps\in(0,\epsp]$  do not depend on $\ka^*$ since they are defined in terms of the {\em unique} extension $J(\eps,.)$ to $\overline{A_{\eps,N}}$.
\end{rem}

\begin{rem} In the case of LIF-like neural models, combination of Proposition \ref{prop: LIF-like} and Corollary \ref{cor: LIF-like} gives $\epsp=\epsm$, i.e., $X_-=\emptyset$ in the proof of Theorem \ref{prop: main}.
%This situation is implicitly included in the above result and its proof as a subcase: one has strict positivity of $Z'(\pp)$ for all $\pp\in[0,1]$, and $X_-=\emptyset$, implying $\epsp=\epsm$.
\end{rem}

\subsection{Synchronization}\label{subsection: main result synchronization}

We are now ready to prove the main theorem on synchronization for a network of $N$ all-to-all coupled oscillators.
The evolution of such a system  with given initial reduced phase vector, may involve phase spaces $S_l$ of dimensions $l$, with $0\leq l\leq n=N-1$, representing configurations after  a number of absorptions have taken place. %We therefore add $l$ in superscript to the definitions below.
Define the set of phase vectors in $S_l$ with no absorption during its time evolution, respectively, the set in $S_l$ that  never synchronizes:
%~\footnote{Comparing to the previous notation, cf.\ \eqref{def-Aeps}, we have $A_\eps=A^n_{\eps}$.}
\beqs
&&A^l_\eps\equiv\{\th\in S_l|h_\eps^i(\th)\in S_l,\;\forall i\in \N\},\\
&&B^l_\eps\equiv\{\th\in S_l|h_\eps^i(\th)\neq 0,\;\forall i\in \N\}.
\eeqs
Referring to the notation in \eqref{def-Aeps}, notice that $A_\eps=A^n_{\eps}$. Also notice that $A^1_\eps=B^1_\eps$. For all $r\geq 1$ and $1\leq k<l$, the set of phase vectors in $B^l_{\eps}$ that survive $r-1$ applications of $h_\eps$ without absorptions, and then get absorbed into $S_k$, is denoted as $B^l_{\eps,r,k}$ and defined for $1< l\leq n$ as
\beqs
&&B^l_{\eps,r,k}\equiv \{\th\in B^l_{\eps}|h_\eps^i(\th)\in S_l,\;i=0,\ldots,r-1,\;h_\eps^r(\th)\in S_k\}.
\eeqs
Furthermore~\footnote{On pp.\ 1656 of \cite{MirStr90} there should be $1\leq k<n$ instead of $1\leq k\leq n$ in the corresponding expression for $B\equiv B^n_\eps$, which is crucial in the proof by induction of the Theorem 3.2 of \cite{MirStr90}; cf.\ also our Proposition \ref{lem: 3-An-Bn}.}
\be\label{B-union}
B^l_\eps=A^l_\eps \;\;\cup\;\bigcup_{\begin{smallmatrix}{1\leq k <l}\\{r\geq 1}\end{smallmatrix}} B^l_{\eps,r,k},  \quad l>1.
\ee

\begin{prop}\label{lem: 3-An-Bn} If $\mu_l(A^l_\eps)=0$ then also $\mu_l(B^l_\eps)=0$ ($1\leq l\leq n$).
\end{prop}

\begin{proof} By strong induction  on $l$ (cf.\ \cite{MirStr90}, Theorem 3.2). Denote the statement of the proposition as $P(l)$. Then $P(1)$ holds since the sets involved are equal. Suppose that $P(k)$ holds for all $k<l$, and that $\mu_l(A^l_\eps)=0$. By \eqref{B-union} it suffices to show that $\mu_l(B^l_{\eps,r,k})=0$ for all $k<l$ and $r\geq 1$. Keeping $k$ fixed we have $h_\eps^{r-1}(B^l_{\eps,r,k})\subset B^l_{\eps,1,k}$ for all $r\geq 1$, where $h_\eps^{r-1}$ acts as a diffeomorphism. Hence, it suffices to show that  $\mu_l(B^l_{\eps,1,k})=0$. By $h_\eps(B^l_{\eps,1,k})\subset B^k_\eps$ and the induction hypothesis we know that $\mu_{k}(B^k_\eps)=0$ and therefore $\mu_k(h_\eps(B^l_{\eps,1,k}))=0$. Here $h_\eps$ acts as $\s_k\circ\tau_{\eps,kk}\circ\pi_{lk}$, cf.\ \eqref{write-taukl}, where $\s_k\circ\tau_{\eps,kk}$ is a diffeomorphism of $S_k$. Hence, $\mu_k(\pi_{lk}(B^l_{\eps,1,k}))=0$, which is only possible if $\mu_l(B^l_{\eps,1,k})=0$.
\end{proof}

Proposition \ref{lem: 3-An-Bn}, for $l=n$ combined with Theorem \ref{prop: main} gives rise to our main result on synchronization.

\begin{thm}\label{thm: main}

Consider an all-to-all network of weakly pulse-coupled identical oscillators with iPRC $Z$ satisfying condition \eqref{Z-cond-p}, then the network will synchronize for almost all initial conditions. %for $\eps>0$ sufficiently small.

%Let $(F,\xon,\xbo)$ be an IF oscillator model, with $F$ and $F'$ continuous on $[\xon,\xbo]$. If the corresponding iPRC $Z$ satisfies condition \eqref{Z-cond-p} [resp., \eqref{Z-cond-m}] then, for sufficiently small $|\eps|$, any excitatory [resp., inhibitory] IF network $(N,F,\xon,\xbo,\eps)$ will achieve synchrony for almost all initial conditions.
\end{thm}

\begin{rem}\label{remark: sum-condition} Based on \eqref{Jeps-epxr}, condition \eqref{Z-cond-p} may be replaced by~\footnote{One has $\overline{S_l}=\{(\ph_1,\ldots,\ph_l)\in \R^l|0\leq\ph_1\leq\ldots\leq\ph_l\leq 1\}$, and the sum in \eqref{sum-cond} should be evaluated for $\ph_0=0$ and for all $(\ph_1,\ldots,\ph_n)\in\overline{S_n}$. Since only differences appear in the arguments of $Z'$, this is equivalent to the evaluation of the sum for all $(\ph_0,\ph_1,\ldots,\ph_n)\in\overline{S_N}$, as seen by writing $\ph_i=\ph_0+(\ph_i-\ph_0),\,i=1,\ldots,n$.}
\be\label{sum-cond}
\sum_{0\leq k<l\leq n}[Z'(\ph_l-\ph_k)+Z'(1-(\ph_l-\ph_k))]>0,\qquad \forall\; (\ph_0,\ph_1,\ldots,\ph_n)\in\overline{S_N}.
%\ph_0=0,\quad \forall \;0\leq \ph_1\leq\ldots\leq\ph_n\leq 1
%\ph=(\ph_1,\ldots,\ph_n)\in \overline{S_n}.
\ee
For fixed $n$, this condition on $Z'$ may be less stringent than \eqref{Z-cond-p}.
%but is less workable.
However, during the synchronization process it needs to be evaluated for every $n'\leq n$, where for $n'=1$ the original condition \eqref{Z-cond-p} is recovered. Therefore, \eqref{Z-cond-p} is a natural and dimension-independent condition, guaranteeing synchronization when using a proof technique based on increasing phase volumes as introduced by Mirollo and Strogatz~\cite{MirStr90}.
\end{rem}

\subsection{Generalization to arbitrary connectivity and PTC's}\label{subsection: arbitrary}

%\subsubsection{Absorption}\label{subsection: arbitrary absorption}

We point out how our results generalize if the pulse-coupling is no longer assumed to be global and homogeneous
(cf.\ Sec.\ \ref{subsec: network}). We still consider $N$ pulse-coupled oscillators with phases $\pp_\a$  and phase dynamics \eqref{phasedyn}.
However, the oscillators are now integrated in an arbitrarily connected network, and have PTC's possibly depending on firing and receiving oscillators.

In this setting, pulse-like interactions between oscillators are modeled by taking, for every $\a$ and $\b$ with $\b\neq \a$, a function $\ka_{\eps,\a\b}:(0,1)\ra(0,1]$. The phase of oscillator $\b$ is assumed to instantaneously change from $\pp_\b$ to $\ka_{\eps,\a\b}(\pp_\b)$ when oscillator $\a$ fires.
%{\bf The pulse strength is modeled as a function $f_{\a\b}$ of the parameter $\eps\in[0,\epsm]$, with $f_{\a\b}(0)=0$}.
For every $\a$ and $\b$, $\b\neq \a$, either $\ka_{\eps,\a\b}$ is the identity map for all $\eps$ (modeling the absence of influence of $\a$ on $\b$), or the collection $\{\ka_{\eps,\a\b}|\eps\in [0,\epsm]\}$ satisfies properties P1-P4 in Sec.\ \ref{subsec: oscillator}. Asymmetry, i.e., $\ka_{\eps,\a\b}\neq\ka_{\eps,\b\a}$ for some $\a$ and $\b$, is allowed. Similar to \eqref{def-k} and \eqref{Z-prop}, one defines
%the {\em single oscillator PTC's}
PTC's by $\ka_{\a\b}(\eps,\pp_\b)=\ka_{\eps,\a\b}(\pp_\b)$, and corresponding iPRC's % of $j$ by the firing of $i$
%In the more general setting, pulse-like interactions between oscillators can be modeled by taking, for every $i$ and $j$ with $j\neq i$, a function $\ka_{\eps,ij}:(0,1)\ra(0,1]$. The phase of oscillator $j$ is assumed to instantaneously change from $\pp$ to $\ka_{\eps,ij}(\pp)$ when oscillator $i$ fires. Here the pulse strength is a function of the parameter $\eps\in[0,\epsm]$, and for every $i$ and $j$, $j\neq i$, $\ka_{\eps,ij}$ is either the identity map for all $\eps$ (modeling the absence of influence of $i$ on $j$), or the collection $\{\ka_{\eps,ij}\}_{\eps\in [0,\epsm]}$ satisfies properties 1-4 in Sec.\ \ref{subsec: oscillator}. Asymmetry ($\ka_{\eps,ij}\neq\ka_{\eps,ji}$) is allowed. In analogy with \eqref{def-k} and \eqref{Z-prop}, one defines corresponding PTC's by $\ka_{ij}(\eps,\pp)=\ka_{\eps,ij}(\pp)$, and iPRC's % of $j$ by the firing of $i$
by
\be\label{iprc-general}
Z_{\a\b}=\fpd{\ka_{\a\b}}{\eps}(0,\pp_\b).
\ee
Notice that $Z_{\a\b}=0$ if there is no influence of $\alpha$ on $\beta$.
%$\ka_{\eps,\a\b}$ is the identity map for all $\eps$.
In this general setting, Theorem \ref{prop: main} is extended to:

\begin{thm}\label{prop: main-general} Consider an arbitrary network of weakly pulse-coupled identical oscillators
%, modeled by a set of PTC's $\ka_{\a\b}$ as described above,
with corresponding iPRC's $Z_{\a\b}$.
Suppose that
\be\label{Z-cond-p-general}
Z'_{\a\b}(\pp)+Z'_{\b\a}(1-\pp)>0,\qquad \forall \pp\in [0,1],
\ee
for any couple of oscillators $(\a,\b),\a\neq\b$ with $Z_{\a\b}$ and/or $Z_{\b\a}$ non-zero. Then an absorption occurs for almost all initial conditions.
\end{thm}

An important subcase arises when the coupling is symmetric and all non-trivial PTC's are the same, i.e., for all $\eps\in[0,\epsm]$ and all $\alpha\neq\beta$ one has $\ka_{\eps,\a\b}=\ka_{\eps,\b\a}$, equal to either the identity map or a certain map $\ka_\eps$. If $\ka$ is the PTC corresponding to the collection $\{\ka_{\eps}|\eps\in [0,\epsm]\}$, and if we define $Z(\pp)\equiv\fpd{\ka}{\eps}(0,\pp)$ for all $\pp$, then condition \eqref{Z-cond-p-general} reduces to \eqref{Z-cond-p}. This means that \eqref{Z-cond-p} is a sufficient condition for absorption (for almost all initial conditions) in the case where the coupling is homogeneous but {\em the connectivity is not necessarily all-to-all}.

The main difficulty to prove Theorem \ref{prop: main-general} for arbitrary pulse-coupling
%originate from the fact
is that when a certain oscillator $\a$ fires, the point representing a pulse-receiving oscillator $\b$ may pass the point representing another oscillator $\g$ ($\pp_\b<\pp_\g$ but $\ka_{\a\b}(\pp_\b)>\ka_{\a\g}(\pp_\g)$). To circumvent this difficulty one needs to define a convenient phase space and firing map (different from those in Sec.\ \ref{subsec: phase space}), and adapt the concepts and results of Sec.\ \ref{subsec: absorption}.
%, and generalize lemma \ref{lem: phi-eps}.
The precise mathematical setting will be introduced, and the proof of Theorem \ref{prop: main-general} outlined, in a forthcoming paper~\cite{WyllAey14}.
%, where networks of arbitrarily pulse-coupled oscillators will be discussed extensively.

If one is willing to accept that clusters of temporarily synchronized oscillators never break up, then clusters can be still treated as single oscillators (cf.\ Sec.\ \ref{subsec: network}). This requires appropriate modifications of the original set of single oscillator PTC's, each time an absorption has taken place. If these sets of PTC's continue to satisfy \eqref{Z-cond-p}, then the network will synchronize for almost all initial conditions. This generalization of Theorem \ref{thm: main} depends crucially on the assumption of impossibility for the clusters to break up, as shown explicitly in \cite{WyllAey14}.

\begin{rem} Without loss of generality, one can rank the {\em initial} oscillator phases, $\ph_\a$, in ascending order
%  ($\pp_\a<\pp_\b$ if $\a<\b$).
($0=\ph_1<\ldots<\ph_N$). In line with Sec.\ \ref{subsubsection: observation} and remark \ref{remark: sum-condition} we observe that, in the limit $\eps=0$ and in the `passive' picture, the arc lengths $\ph_\b-\ph_\a$ and $1-(\ph_\b-\ph_\a)$ will occur as arguments of $\ka_{\eps,\a\b}$ and $\ka_{\eps,\b\a}$, respectively (i.e., when $\a$ `fires' and $\beta$ receives the `pulse', and vice versa). Hence, condition \eqref{sum-cond} generalizes to
\be\label{sum-cond-general}
\sum_{1\leq\a<\b\leq N}
Z'_{\a\b}(\ph_\b-\ph_\a)+Z'_{\b\a}(1-(\ph_\b-\ph_\a))>0,\qquad \forall \ph=(\ph_1,\ldots,\ph_N)\in \overline{S_N},
\ee
which guarantees the occurrence of at least one absorption, for almost all initial conditions.
\end{rem}

\subsection{Inhibitory coupling}\label{subsection: inhibitory}

The above results are readily adapted to the case where the coupling is inhibitory, $\eps<0$.
%One now considers a collection of PTC$_\eps$'s, $\{\ka_\eps:(0,1)\ra[0,1)\}_{\eps\in[\epsu,0]}$ for some lower bound $\epsu<0$, satisfying the following properties:
%\begin{enumerate}
%\item $\ka_0(\pp)=\pp$, for all $\pp\in(0,1)$;
%\item $\ka_{\epsu}(\pp)=0$, for all $\pp\in(0,1)$;
%\item for any $\eps\in(\epsu,0)$ the map $\ka_\eps$ is phase-receding %has a graph  above the first bisector
%($\ka_\eps(\pp)<\pp$),
%strictly increases on an interval $(E(\eps),1)$, with $0<E(\eps)\leq 1$, and satisfies $\ka_\eps(\pp)=0$ for all $\pp< E(\eps)$. The map $E:(\epsu,0)\ra(0,1]$ is continuous, with limits $1$ and $0$ if $\eps$ approaches $\epsu$ and $0$, respectively.
%{\bf misschien nog iets over tweede afgeleiden en het feit dat afgeleide naar epsilon enkel in 0}
%\item the PTC
%\begin{equation}\label{def-k-inhib}
%\ka:\quad \kaset\ra (0,1);
%\quad (\eps,\pp)\mapsto \ka(\eps,\pp)\equiv \ka_\eps(\pp),\qquad \kaset\equiv \bigcup_{\eps\in(\epsu,0)}\{\eps\}\times (E(\eps),1),
%\end{equation}
%as well as all its partial derivatives up to second order are continuous on $\kaset$, and can be (uniquely) extended to continuous maps on $\kasetc$.
%\end{enumerate}
Secs.\ \ref{subsec: oscillator} up to and including \ref{subsubsection: observation}, as well as Sec.\ \ref{subsection: main result synchronization}, remain formally unchanged. The statements and proofs of
%lemma \ref{lem: phi-eps} and
the theorems can be straightforwardly modified. Theorems \ref{prop: main} and \ref{thm: main} are replaced by:

\begin{thm}\label{prop: main-inhib} Consider an all-to-all network of identical oscillators with inhibitory pulse-coupling, modeled by a PTC $\ka$ satisfying the above properties. If the corresponding iPRC\, $Z$, defined by \eqref{Z-prop}, satisfies
\be\label{Z-cond-p-inhib}
Z'(\pp)+Z'(1-\pp)<0,\qquad \forall \pp\in[0,1],
\ee
then there exists $\epst<0$ such that, for any $\eps\in[\epst,0)$, $\mu_n(A_\eps)=0$ and the network will synchronize for almost all initial conditions.
\end{thm}

Inversion of the inequality signs applies also to \eqref{sum-cond}, \eqref{Z-cond-p-general} and \eqref{sum-cond-general}.

\section{Synchronization of a network of QIF oscillators}\label{sec: QIF}

In this section we check when \eqref{Z-cond-p} is satisfied for neural IF networks of the quadratic-integrate-and-fire type (QIF) with excitatory coupling. Similar results may be obtained in the inhibitory case, {\em mutatis mutandis}.

A {\em quadratic integrate-and-fire (QIF)} oscillator is characterized by a triple $(F,\xon,\xbo)$, with
\be
F(x)=S+x^2,\qquad S\in\R_{>0},\qquad \forall x\in[\xon,\xbo].
\ee
There is no loss of generality in assuming $S=1$, taking a new state $x/S^{1/2}$ and time $tS^{1/2}$, but we keep the original variables. Notice that
\be
\frac{F'}{F}(x)=\frac{2x}{S+x^2}.
\ee
It follows that if $\xbo<0$ then $F'/F(x)<0$ for any $x\in[\xon,\xbo]$, from which $Z'(\pp)>0$ for any $\pp$, see \eqref{Z-cond-alt}. Condition \eqref{Z-cond-p} is automatically satisfied and guarantees  synchronization; this reminds us of the condition put forward in  \cite{MirStr90} for the LIF case. On the other hand, if $\xon\geq 0$ then $F'/F(x)\geq 0$ for any $x\in[\xon,\xbo]$; in other words: $Z'(\pp)\leq 0$ for any $\pp$ and condition \eqref{Z-cond-p} is never satisfied. From now on we  assume $\xon<0\leq \xbo$.
From \eqref{period}, \eqref{state-phase} one obtains %by straightforward calculation that
\bes
x=f(\pp)=\sqrt{S}\tan(\pp\beta+(1-\pp)\alpha),\qquad T=\frac{1}{\sqrt{S}}(\b-\a),
\ees
where
\[
\alpha\equiv \arctan\left(\frac{\xon}{\sqrt{S}}\right),\qquad \beta\equiv \arctan\left(\frac{\xbo}{\sqrt{S}}\right),
\]
with $-\frac{\pi}{2}<\a<\b<\frac{\pi}{2}$. From \eqref{def-iPRC} one obtains
\bes
Z(\pp)=\frac{1}{ST}\cos^2(\pp\beta+(1-\pp)\alpha)=\frac{1}{2ST}\left[1+\cos(2\pp\beta+2(1-\pp)\alpha)\right].
\ees
%By a straightforward calculation one gets that
%Hence,
%\be
%Z'(\pp)=-\frac{1}{\sqrt{S}}\sin(2\pp\beta+2(1-\pp)\alpha)
%\ee
%and the sufficient condition \eqref{Z-cond} thus becomes
The sufficient condition \eqref{Z-cond-p}  becomes
\be\label{sufficient-QIF}
%m(\pp)\equiv
Z'(\pp)+Z'(1-\pp)=-\frac{2}{\sqrt{S}}\sin(\a+\b)\cos((2\pp-1)(\b-\a))>0,\qquad \forall \pp\in[0,1].
\ee
Since $-\frac{\pi}{2}<\a<\b<\frac{\pi}{2}$ %and $2\pp-1\in[-1,1]$
it follows that  $-\pi<\a+\b<\pi$ and $0<\b-\a<\pi$. Taking  $\pp=\frac{1}{2}$ in \eqref{sufficient-QIF} leads to $\sin(\a+\b)<0$ or $\a+\b<0$. Condition \eqref{sufficient-QIF} is reduced to
\[
\cos((2\pp-1)(\b-\a))>0,\qquad \forall \pp\in[0,1]
\]
which leads to $\b-\a<\frac{\pi}{2}$.
To conclude, the sufficient condition \eqref{sufficient-QIF} is equivalent with
\be\label{cond-ab}
\a+\b<0,\qquad \b-\a<\frac{\pi}{2},
\ee
or  written in model parameters
\be\label{cond-xon-xbo}
\xon+\xbo<0,\qquad S+\xon\xbo>0.
\ee
%This is automatically satisfied if $\b\leq 0\lra \xbo\leq 0$, and never satisfied if $\a\geq 0\lra \xon\geq 0$. In these cases, **** ****For $\xon<0<\xbo$ one finds that \eqref{cond-ab} is equivalent to
%\be
%\xbo<\min\left(|\xon|,\frac{S}{|\xon|}\right)\quad\lra\quad
%\xbo<\begin{cases}
%|\xon|,&|\xon|<\sqrt{S},\\
%\frac{S}{\xon},&|\xon|>\sqrt{S},
%\end{cases}
%\qquad (\xon<0<\xbo).
%\ee
Notice that this implies $0<\xbo<\sqrt{S}$, and we find that for any given such $\xbo$,
%equivalently we find that
condition \eqref{Z-cond-p} is satisfied iff $-\frac{S}{\xbo}<\xon<-\xbo$. If $\xbo\leq 0$ then \eqref{cond-xon-xbo} is automatically satisfied.
%either $\xbo\leq 0$ or
%\be\label{cond-QIF}
%\xbo\leq 0\qquad\textrm{or}\qquad \left(0<\xbo<\sqrt{S},\quad -\frac{S}{\xbo}<\xon<-\xbo\right).
%\left(\xon<0<\xbo<\sqrt{S},\quad \xbo <|\xon|<\frac{S}{\xon}\right).
%\ee
As an application of our main theorem \ref{thm: main} we conclude:
\begin{cor} Let $(F,\xon,\xbo,\eps)$ be a QIF oscillator network, with $F(x)=S+x^2,\;x\in[\xon,\xbo]$. Then, if \eqref{cond-xon-xbo} holds and $\eps$ is small enough, the network will synchronize for almost all initial conditions.
\end{cor}

The second condition in \eqref{cond-xon-xbo} is technical: simulations suggest that it is not needed for absorption or synchronization. It is a consequence of the technique used in proving absorption, which is based on volume increasing properties of the dynamics of the interconnected set of oscillators.

\begin{comment}
\begin{figure}[h!]
\begin{center}
\includegraphics[width=\textwidth]{./figuur_QIF.pdf}
\end{center}
\caption{The dashed region}
\label{fig:QIF}
\end{figure}
\end{comment}

\begin{figure}[t]
    \begin{center}
        \subfigure[]{
            \label{fig: QIF-a}
            \includegraphics[width=0.4\linewidth]{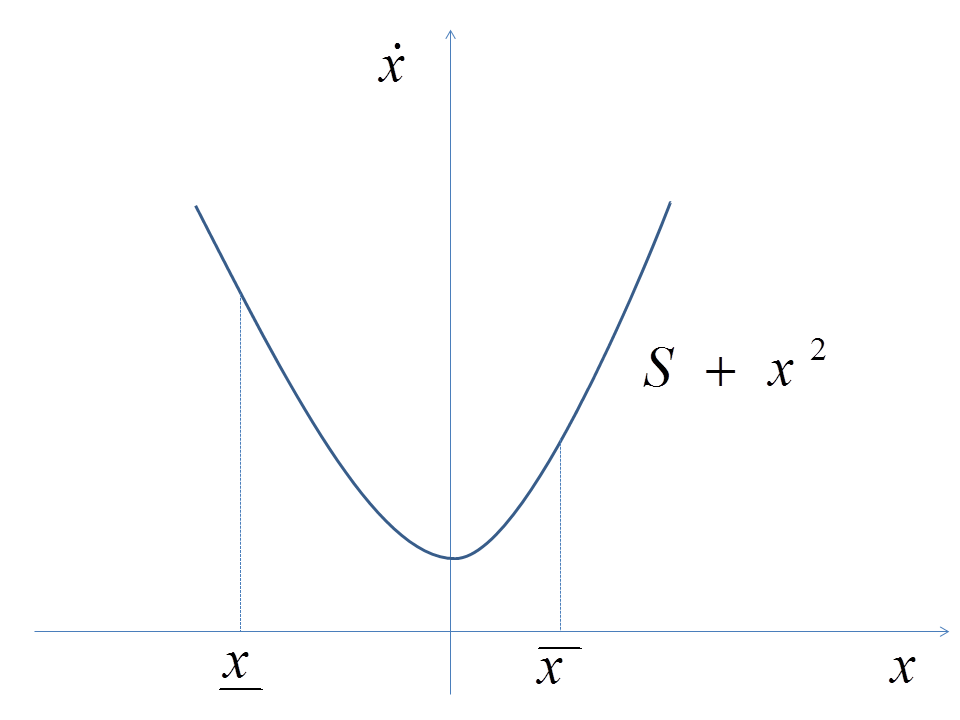}
        }
        \subfigure[]{
           \label{fig: QIF-b}
           \includegraphics[width=0.4\linewidth]{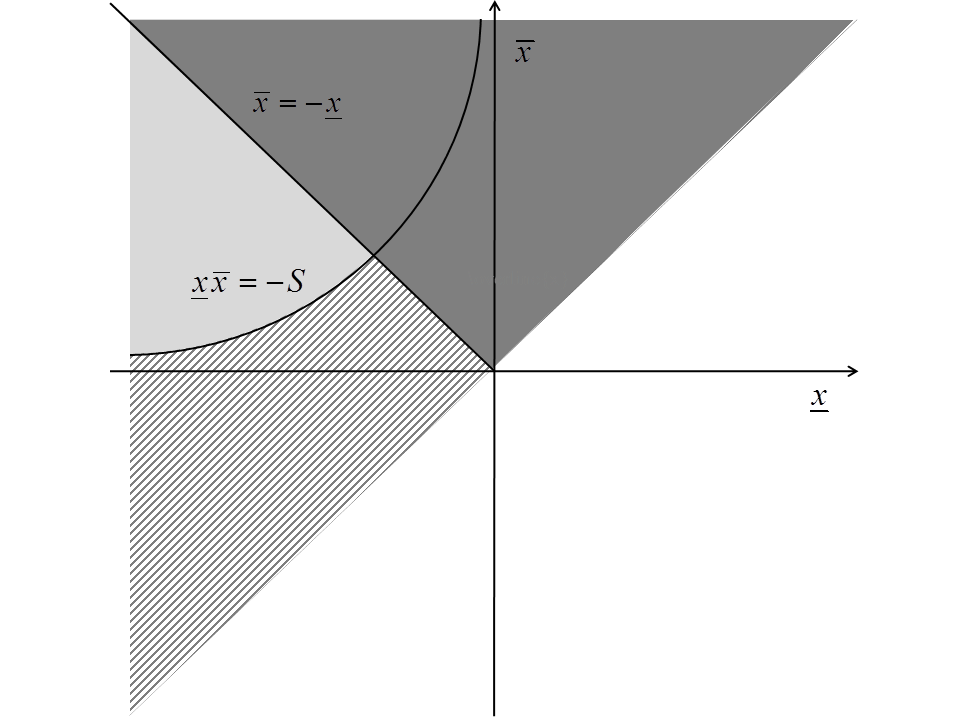}
        }
    \end{center}
    \caption{(a) A QIF oscillator model $(F,\xon,\xbo)$, with $F(x)=S+x^2$. (b) Condition \eqref{Z-cond-p} for a QIF model in function of $\xon$ and $\xbo$ (with $\xbo>\xon$ and for given $S$). In the dark gray region $(\xon+\xbo\geq 0)$ condition \eqref{Z-cond-p} is nowhere satisfied, whereas it is satisfied for some (but not all) values of $\varphi$ in the light gray region. The shaded region corresponds to \eqref{cond-xon-xbo}: for these parameter values  synchronization is achieved for almost all initial conditions.}
    \label{fig: QIF}
\end{figure}

%\pagebreak

  \section{Summary and Discussion}\label{sec: discussion}

In this paper we propose a sufficient condition, phrased in terms of the infinitesimal phase response curve (iPRC), for synchronization of fully coupled networks of weakly pulse-coupled oscillators. A description in terms of the iPRC is preferred over the  formulations featuring the phase-state map commonly used in the literature on  neural integrate-and-fire network models, allowing for more general pulse-coupled networks.

%This condition is formulated in terms , which we introduced in relation with a more general notion of phase transition curve than is usual in the literature on neural integrate-and-fire network models.

The  condition on the iPRC is conceptually different from conditions featuring in the literature (see e.g.~\cite{MirStr90}) which have a local flavor. Our condition  is of a global nature, reflecting the presence of the circle as the space where the oscillators live on. Our results are  independent of the number of oscillators involved. They derive from exploring the uncoupled dynamics ($\epsilon=0$) and pushing this to a condition for absorption for small $\epsilon$.
Describing the dynamics of the network of oscillators by the return map, the condition implies that, for weak pulse coupling, the phase volume increases {\em everywhere} in the phase space  at each application of the return map. Our results include and extend the synchronization results of Mirollo and Strogatz (MS)~\cite{MirStr90}, who considered the particular case of a network of neural oscillators with a strictly increasing iPRC, or equivalently with a concave down phase-state map, and showed that this ensures synchronization.
%we lifted information available from the limit case $\eps=0$ to weak pulse coupling, with small but finite $\epsilon>0$ (or $\epsilon<0$ in the inhibitory case). Crucial observation, passive picture -> complementarity $(\pp,1-\pp)$ essential here. leads to a workable, single condition independent of the number of clusters, if one sticks to a global, homogeneous, non-additive pulse coupling model.

The main focus is on all-to-all and homogeneous coupling. However, our conditions enforce absorption  in not fully and heterogeneously coupled networks, illustrating robustness of the absorption phenomenon against network changes. This extension of our approach, enabling us  to consider more general networks, and also coping with particular time delays, will be discussed in more detail in a follow-up paper~\cite{WyllAey14}.

To illustrate the wide applicability of our theorems, we show that synchronization is secured (for almost all initial conditions) for networks of quadratic integrate-and-fire (QIF) neural  cells with a negative sum of the thresholds (see also~\cite{MauSacSep12,Mauroythesis})
%phase-state map that is concave down in the mean~\cite{Mauroythesis}
 and satisfying an extra condition on the model parameters.
Computer simulations suggest the  conjecture  that this extra condition is redundant~\cite{Mauroythesis}. If true this  would illustrate that the volume-increasing approach  has its limitations; stronger results on synchronization may be obtained for networks of specific classes of oscillators by adopting an ad hoc approach.
%merely an effect of the phase volume technique and superfluous.
%**************An analytical proof of this conjecture will be the subject of a future paper. {\bf Spreken we hier al over een titel?}
%*****best nog iets bij over grote epsilon en synchronisatie??????
%%For non-negligible delay times we must replace $\ka(\pp)$ by $\ka(\pp+\omega \Delta t_{\textrm{tr}})+\pp_0$, where $\pp_0=\omega(\Delta t_{\textrm{tr}}-\Delta t_{\textrm{r}})$.
%{\bf weglaten However, as illustrated in \cite{Mauroyetal10} one should be careful with conclusions from numerical simulations, and a further analytic treatment is desirable.}

\section {Acknowledgment}
The authors have had stimulating discussions with Jonathan Rogge and Hans Vernaeve.  This paper presents research results of the Belgian Network DYSCO (Dynamical Systems, Control, and Optimization), funded by the Interuniversity Attraction Poles Programme initiated by the Belgian Science Policy Office. Lode Wylleman has been financially supported by the DYSCO network.

\end{document}